# Emergence of Urban Heat Traps from the Intersection of Human Mobility and Heat Hazard Exposure in Cities

Xinke Huang[1]*, Yuqin Jiang[2], Ali Mostafavi[3]

[1] Ph.D. Student, Zachry Department of Civil and Environmental Engineering, Texas A&M University, College Station, United States; e-mail: adahuang@tamu.edu

[2] Postdoctoral Researcher, Urban Resilience.AI Lab Zachry Department of Civil and Environmental Engineering, Texas A&M University, College Station, United States; e-mail: yuqinjiang@tamu.edu

[3] Associate Professor, Urban Resilience.AI Lab Zachry Department of Civil and Environmental Engineering, Texas A&M University, College Station, United States; e-mail: amostafavi@civil.tamu.edu

**Abstract**

Understanding the relationship between spatial structures of cities and environmental hazard exposures (such as urban heat) is essential for urban health and sustainability planning. However, a critical knowledge gap exists in terms of the extent to which socio-spatial networks shaped by human mobility exacerbate or alleviate urban heat exposures of populations in cities. In this study, we utilize location-based data to construct human mobility networks in twenty metropolitan areas in the U.S. The human mobility networks are analyzed in conjunction with the urban heat characteristics of spatial areas. We identify areas with high and low urban heat exposure and evaluate visitation patterns of populations residing in high and low urban heat areas to other spatial areas with similar and dissimilar urban heat exposure. The results reveal the presence of urban heat traps in the majority of the studied metropolitan areas in which populations residing in high heat exposure areas primarily visit areas with high heat exposure. The results also show a small percentage of human mobility to produce urban heat escalate (visitations from low heat areas to high heat areas) and heat escapes (movements from high heat areas to low heat areas). The findings from this study provide a better understanding of urban heat exposure in cities based on patterns of human mobility. These finding contribute to a broader understanding of the intersection of human network dynamics and environmental hazard exposures in cities to inform more integrated urban design and planning to promote health and sustainability.

## Introduction

The characterization of the spatial environmental hazards in cities is essential for urban sustainability and health plans and policies (Shen et al., 2011, Seo et al., 2019, Hunter et al., 2019). Among all the environmental hazards, heat is one of the major hazards. Damages of heat include increased mortality and morbidity due to extremely high air temperatures (Kim & Brown, 2021), stronger heat-related health threats in urban areas (Li et al., 2016), and increased energy consumption (Xie et al., 2019). However, comparing to other environmental hazards, such as air pollution, urban heat did not draw enough attention in the existing literature (Bao et al., 2022, Glencross et al., 2020, Venter et al., 2020). Within the studies of urban heat, limited attentions were paid to human network dynamics that could expand the reach of environmental hazard exposures (Coccia, 2020). Current heat-related studies mostly focused on index-based, which is an isolated measurement of individual locations (Andrade & Szlafsztein, 2018, Jha & Gundimeda, 2019, Orioli et al., 2019). Research gap exists in terms of how to understand the spatial distribution of urban heat and people's respond to the heat from a network-based perspective. In particular, human mobility shapes the spatial structures of cities and could extend the reach of environmental hazards beyond hazard hotspots. In a recent study, Fan et al. (2022) examined the intersection of human mobility and air pollution exposure and found that human mobility expands the reach of air pollution exposure. This study highlights the significance of characterizing environmental hazard exposures based on considering human mobility networks in cities (Fan et al., 2022). In the context of urban heat exposure, Yin et al. (2021) proposed a dynamic urban thermal exposure index to account for human mobility in specifying urban heat exposure. While the index-based approach proposed by Yin et al. (2021) captures mobility-based heat exposure, it does not capture fundamental properties arising at the intersection of human mobility and spatial heat exposure that extend or alleviate heat exposure. Recognizing this gap, in this paper, we define and examine three properties at the intersection of urban heat and human mobility (Figure 1): (1) *heat traps*: in which populations residing in high heat areas visit other high heat areas; (2) *heat escapes*: in which populations residing in high heat areas visit low heat areas; and (3) *heat escalates*: in which populations residing in low heat areas visit high heat areas. In fact, these properties are emergent properties arising from the intersection of human mobility networks and the spatial distribution of heat hazards in cities. Accordingly, the study aims to address the following research questions: to what extent human mobility would exacerbate urban heat exposure (prominence of heat traps), alleviate heat exposure (heat escapes), or expand the reach of heat exposure (heat escalates)? To address these questions, we utilize aggregated and anonymized location-based data to construct the human mobility network (origin-destination network in which origin is the home census tracts of trips and destination is the visitation census tract of trips) for twenty metropolitan areas in the U.S. to examine the proportion of trips from high heat areas to other high heat areas and low heat areas. Accordingly, we analyze the prominence of heat traps, escapes, and escalates across different cities to evaluate cross-city similarities and differences.

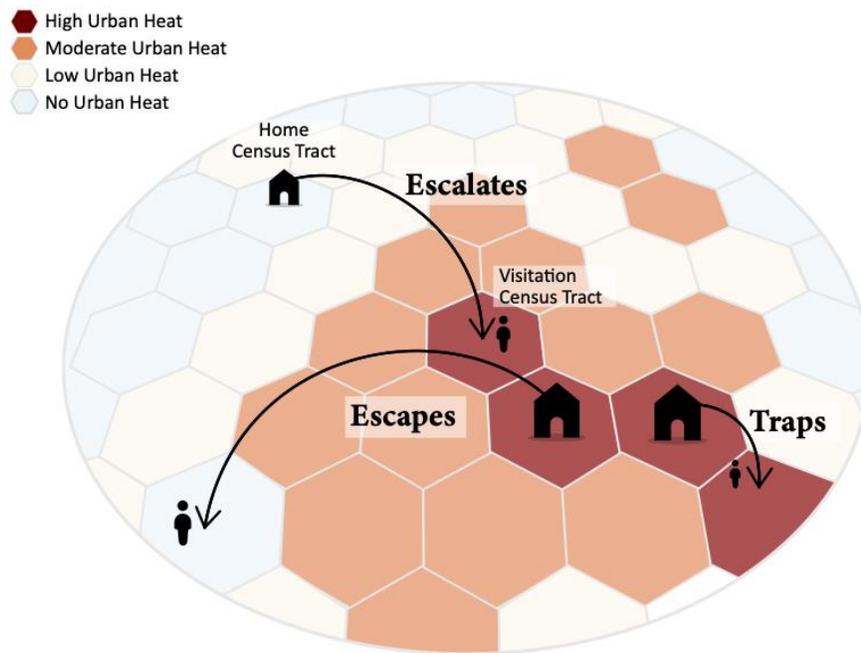

**Figure 1**. Conceptual representation of urban heat traps, escalates, and escapes arising from the intersection of human mobility and heat exposure.

## Background

Urban heat (UH), or the urban heat island effect, refers to the phenomenon where urban areas have higher temperatures than surrounding rural areas due to the heat generated by human activity and the lack of vegetation to absorb that heat. To understand and mitigate UH effect, researchers have identified multiple factors. For example, some studies found that tree density is correlated with UH (Ziter et al., 2019, Rahman et al., 2020, Morabito et al., 2021), that high tree density potentially decreases urban heat phenomenon. Transportation is another factor, less movement of transportation can reduce the extent of changes in temperatures in urban areas (Hu et al., 2019, Ali et al., 2021, Angelevska et al., 2021). Moreover, population density also contributes to the urban heat effect, population loss can have a mitigating effect on the UH effect (Zhou et al., 2018, Manoli et al., 2019, Peng et al., 2022). However, those studies focused on examining a single factor with UH, that they ignored the ability for human to adjust living environment by moving to different locations.

Human mobility datasets have been widely used in multiple hazards, including hurricane (Li et al., 2020, Rajput et al., 2020, Dargin et al., 2021, Li & Mostafavi, 2022, Paradkar et al., 2022), flooding (Esparza et al., 2022, Farahmand et al., 2022a, Farahmand et al., 2022b, Mostafavi & Yuan, 2022, Ridha et al., 2022, Yuan et al., 2022a, Yuan et al., 2022b), and infectious diseases (Fan et al., 2021, Ma et al., 2022, Li et al., 2021, Rajput et al., 2022). These studies have found human mobility data was useful to understand people's reaction to hazards (Lai et al., 2019). For example, when hurricane comes, people in the similar social media networks were likely to make the same evacuation decisions (Jiang et al., 2019). During the COVID 19 pandemic, the confirmed cases were found highly correlated with human mobility that places with higher activities had more covid cases (Coleman et al., 2022, Huang et al., 2020). These studies have recognized that people can successfully change the level of hazard exposure by moving to a different location.

The majority of human mobility and hazard studies have focused on the relationship between human mobility patterns and the likelihood of exposure to natural hazards, infectious diseases, and environmental pollutants. However, the current literature does not adequately investigate the relationship between human mobility and urban heat (Smith et al., 2019). In this context, mobility can play a significant role in determining the likelihood of exposure to urban heat. Therefore, understanding the relationship between human mobility and UH can be useful in developing strategies to reduce the impact of urban heat on individuals and communities, which is the focus of this study.

## Data Description

### Study Context

We collected mobility data in February 2020 in twenty metropolitan areas (Table 1) in the U.S. to construct human mobility networks. The rationale for selecting February 2020 is that it was just before the start of the COVID-19 pandemic, and the patterns of human mobility would represent the standard patterns of mobility.

**Table 1**. Metropolitan Areas

|   | **Metropolitan Areas** | **State** |
| --- | --- | --- |
| 1 | Phoenix | Arizona |
| 2 | Los Angeles | California |
| 3 | Denver | Colorado |
| 4 | Washington DC | District of Columbia |
| 5 | Orlando | Florida |
| 6 | Miami | Florida |
| 7 | Atlanta | Georgia |
| 8 | Chicago | Illinois |
| 9 | Boston | Massachusetts |
| 10 | Detroit | Michigan |
| 11 | Minneapolis | Minnesota |
| 12 | Rochester | New York |
| 13 | Columbus | Ohio |
| 14 | Portland | Oregon |
| 15 | Pittsburgh | Pennsylvania |
| 16 | Philadelphia | Pennsylvania |
| 17 | Memphis | Tennessee |
| 18 | Houston | Texas |
| 19 | Dallas | Texas |
| 20 | Seattle | Washington |

**Data sources**

The heat exposure data were obtained from the United States Surface Urban Heat Island database (Chakraborty et al., 2020). For all census tracts in the U.S. urbanized regions, this dataset includes yearly, summer, and winter daytime and nighttime Land Surface Temperature (LST), Digital Elevation Model (DEM), and Normalized Difference Vegetation Index (NDVI) data, as well as the mean values for the whole urbanized area (Chakraborty et al., 2020). The UHI dataset in the urbanized areas was determined by remote sensing data, such as Moderate Resolution

Imaging Spectroradiometer (MODIS) and Global Multi-Resolution Terrain Elevation Data (GMTED), including 55,871 census tracts organized into 497 urbanized areas, covering roughly 78 percent of the population of the United States (Chakraborty et al., 2020). Our study used the mean values for Urban Heat Islands (UHIs) as the measurement of UH for the chosen metropolitan areas. We used quantile breaks to split the UHI data into three clusters and defined them as low UHI area, median UHI area, and high UHI area, respectively.

The location-based data is provided by Spectus (formerly known as Cuebiq), a platform for mobility data. Spectus provides privacy-protected and anonymized location datasets by collecting data from smart devices whose owners have authorized location data collection. Spectus constructs its geo-location dataset by collaborating with application developers to collect high-resolution datasets using Bluetooth, GPS, WiFi, and IoT signals. Each day, more than one hundred data points are gathered for each anonymous user, allowing a more accurate understanding of human movement and visitation patterns. Spectus collects data on around 15 million daily active users in the U.S. High privacy policy standards are set to enable data collection and use of data responsibly and ethically. Users are allowed to opt out of location sharing at any stage, and all information is obtained transparently with consent. All data provided by Spectus is de-identified to ensure anonymity and endures further privacy improvements, such as removing sensitive points of interest and obscuring dwelling locations at the census block group level. In addition to delivering location-based data at the device level, Spectus aggregates data using artificial intelligence and machine learning algorithms. By offering access to an auditable and on-premise sandbox environment, Spectus' platform for responsible data sharing allows us to query anonymized, aggregated, and privacy-enhanced data (Wang et al., 2019). In this study, we used one of the aggregated datasets from Spectus, the Device Location database, to determine the Census tracts of devices' home locations. The Device Location table includes a timestamp, a privacy-compliant device ID, and geoinformation at the device level. To evaluate UH exposure, we used population activity in February 2020, which reflects a steady-state period with no events that could affect population activity and movement.

## Methods

### Mobility network from the home Census tract to the visitation Census tract

Data processing consisted of utilizing Spectus data to construct the human mobility network models. Specifically, it involves two steps. First step is to identify each device's home tract. The second step is to construct the mobility networks. A device's home tract was determined based on its dwell times, as Spectus provides dwell time at each location.

By using unique identifiers for each device, Spectus can collect each visitor's destination tract and aggregate the number of visits from one tract to another tract. Accordingly, we construct the monthly mobility network model of each city, which captures the number of visits from home tracts to visitation tracts. In this network, each node is a tract and the links are number of trips observed between each pair of tracts.

### The Ratio of Urban Heat Traps, Escalates, and Escapes

In each metropolitan area, we used quantile breaks dividing Census tracts into low UH areas, median UH areas, and high UH areas. In this study, we only considered low and high UH areas. We aggregated human mobility dataset to summarize the number of trips between low and high

UH areas. As noted earlier, we define heat traps as high UH areas whose populations visit places in other high UH areas. Similarly, heat escalates are low UH exposure areas whose populations visit places in high UH areas. And heat escapes are high UH exposure areas whose populations visit places in low UH areas. The ratio of UH traps, escalates, and escapes of each tract is calculated by summing the trips in each category (high to high, low to high, and high to low, respectively) and dividing by the total trips associated with each home tract. The ratio of heat escalates is computed using Equation 1:

$$RLow_{i,j} = \frac{Census\ TractD_{high_{i,j}}}{TOT_i} \qquad (1)$$

where, $RLow_{i,j}$ refers to the ratio of trips visiting from low UH tract *i* to high UH *j*, $Census\ TractD_{high_{i,j}}$ refers to the total number of trips from low UH tract *i* to high UH tract *j*, and $TOT_i$ refers to the total number of trips starting from origin tract *i*. Similarly, the ratio of trips visiting from high UH tract to low UH tract and the ratio of trips visiting from high UH tract to high UH tracts are computed using Equations 2 and 3, respectively:

$$RHigh_{i,j} = \frac{Census\ TractD_{low_{i,j}}}{TOT_i} \qquad (2)$$

$$RHigh_{i,j} = \frac{Census\ TractD_{high_{i,j}}}{TOT_i} \qquad (3)$$

**Classifying Cities**

For each metropolitan area, we first calculated the total number of tracts in high and low UH exposures based on the UH dataset. Then, we classified cities as heat traps, heat escalates, and heat escapes based on the percentage of trips in each category. If more than half of trips in the city were heat trap type, we classified this cities as urban heat traps. Similarly, if the city has more than half heat escalate trips or heat escape trips, the city was classified as a heat escalate city or heat escape city, respectively.

## Results

### Patterns across Cities

Table 2 presents the list of metropolitan areas, and their percentage of trips in each category (i.e., high to high, low to high, and high to low). The high UH and low UH percentages divide the total number of census tracts by the number of census tracts in high UH areas and low UH areas. Note that the total number of census tracts with trips from high to low and with trips from high to high is the same, but the ratio of trips visiting from high UHI census tract $i$ to low UHI census tract $j$ are significantly different (Equation (2) and (3)). The metropolitan classifications are based on the percentage of low-to-high trips, high-to-low trips, and high-to-high trips, as stated in the previous section.

**Table 2**. Metropolitan areas with the total number of census tracts (CT), different UH visiting patterns count and percentage, and classification of the metropolitan areas.

| MSA | Total # of CT | Total # CT in high UHI areas | High UHI % | Total # CT in low UHI areas | Low UHI % | Total # CT with trips from low to high | Low to high trips % | Total # CT with trips from high to low | Total # of CT with trips from high to high | High to low % | High to high % | Classifications |
|---|---|---|---|---|---|---|---|---|---|---|---|---|
| Atlanta, GA | 885 | 186 | 0.21 | 280 | 0.32 | 37 | 0.13 | 75 | 75 | 0.4 | 0.4 | trap |
| Boston, MA | 947 | 343 | 0.36 | 264 | 0.28 | 10 | 0.04 | 128 | 133 | 0.37 | 0.37 | trap |
| Chicago, IL | 1,923 | 945 | 0.49 | 301 | 0.16 | 168 | 0.56 | 739 | 739 | 0.78 | 0.78 | trap |
| Columbus, OH | 340 | 155 | 0.46 | 67 | 0.2 | 21 | 0.31 | 155 | 155 | 1 | 1 | escalate & trap |
| Dallas. TX | 1122 | 575 | 0.51 | 110 | 0.1 | 42 | 0.38 | 282 | 282 | 0.49 | 0.49 | trap & escape |
| DC | 179 | 56 | 0.31 | 49 | 0.27 | 49 | 1 | 56 | 56 | 1 | 1 | escalate & trap |
| Denver, CO | 581 | 218 | 0.38 | 98 | 0.17 | 5 | 0.05 | 96 | 96 | 0.44 | 0.44 | escape |
| Detroit, MI | 1,158 | 658 | 0.57 | 183 | 0.16 | 31 | 0.17 | 408 | 408 | 0.62 | 0.62 | trap |
| Houston, TX | 908 | 434 | 0.48 | 130 | 0.14 | 86 | 0.66 | 402 | 402 | 0.93 | 0.93 | trap |
| Los Angeles, CA | 2788 | 1462 | 0.52 | 351 | 0.13 | 284 | 0.81 | 1179 | 1179 | 0.81 | 0.81 | escalate & trap |
| Memphis, TN | 221 | 93 | 0.42 | 51 | 0.23 | 50 | 0.98 | 93 | 93 | 1 | 1 | escalate & trap |
| Miami, FL | 1206 | 514 | 0.43 | 232 | 0.19 | 97 | 0.42 | 291 | 291 | 0.57 | 0.57 | escalate & trap |

| City | | | | | | | | | | | | |
|---|---|---|---|---|---|---|---|---|---|---|---|---|
| Minneapolis, MN | 683 | 306 | 0.45 | 120 | 0.18 | 31 | 0.26 | 170 | 170 | 0.56 | 0.56 | trap |
| Orlando, FL | 299 | 99 | 0.33 | 58 | 0.19 | 26 | 0.45 | 88 | 88 | 0.89 | 0.89 | escalate & trap |
| Philadelphia, PA | 968 | 279 | 0.29 | 274 | 0.28 | 20 | 0.07 | 238 | 238 | 0.85 | 0.85 | trap |
| Phoenix, AZ | 893 | 327 | 0.37 | 165 | 0.18 | 157 | 0.95 | 326 | 326 | 1 | 1 | escalate & trap |
| Pittsburgh, PA | 599 | 190 | 0.32 | 203 | 0.34 | 99 | 0.49 | 168 | 169 | 0.88 | 0.88 | escalate & trap |
| Portland, OR | 334 | 178 | 0.53 | 58 | 0.17 | 25 | 0.43 | 106 | 106 | 0.6 | 0.6 | escalate & trap |
| Rochester, NY | 206 | 102 | .0.50 | 17 | 0.08 | 12 | 0.71 | 102 | 102 | 1 | 1 | escalate & trap |
| Seattle, WA | 660 | 200 | 0.3 | 155 | 0.23 | 97 | 0.63 | 120 | 120 | 0.6 | 0.6 | escalate & trap |

**Cities with High Urban Heat Traps**

The Los Angeles metropolitan area shows significant urban heat traps. Figure 2A maps the UH in Los Angeles. Three orange shades represent three levels of UH. The darker the shade is, the more severe UH was observed. The metropolitan area has 13 percent of the tracts in low UH areas, mainly located on the north and east, while 52 percent of the metropolitan area is in high UH areas (dark orange). Figure 2B to 2D shows the ratio of trips between low UH tracts and high UH tracts, which break into four categories for better visualization. Light blue shows a low ratio of trips, and dark blue shows a high ratio of trips. All the following figures are presented in the same plot format as Figure 2A and 2B to 2D.

Figure 2B shows the ratio of trips visiting from low UH tracts to high UH tracts. A high ratio of low-to-high trips from 0.22 to 0.35 occurs in the north, which means that a significant number of people living in low UH areas are visiting high UH areas in the north. Figure 2D shows the ratio of trips from high UH tracts to low UH tracts with a higher ratio of trips, 0.05 to 0.11, occurring in the northwest and southwest. This means that a relatively high number of people living in high UH areas are visiting low UH areas in the northwest and southwest. Figure 2C shows the ratio of trips visiting from high UH tracts to high UH tracts. 81 percent of all the tracts in high UH areas have trips trapped inside high UH areas with the ratio of trips from 0.30 to 0.92, meaning lots of people suffering UH did not move to relief their UH exposure. These urban heat traps are in the northwest and central of Los Angeles, with an especially high ratio from 0.76 to 0.92 in the central. Figure 2D shows the ratio of trips visiting from high UH areas to low UH areas with ratio of trips from 0 to 0.11.

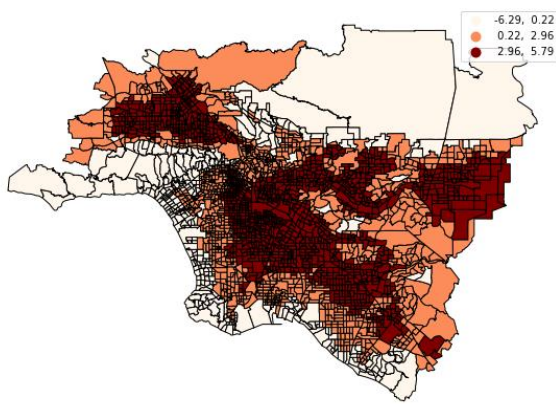

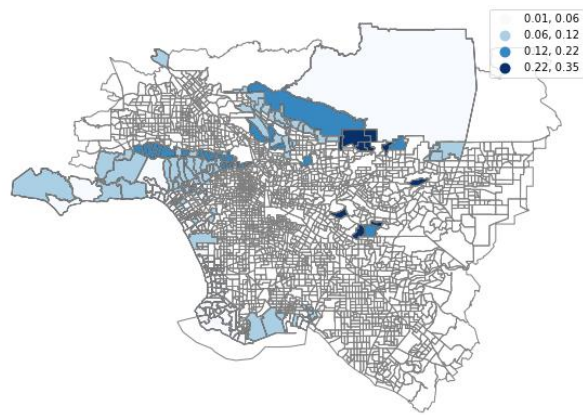

(A) Distribution of urban heat  (B) The ratio of trips from low UH to high UH

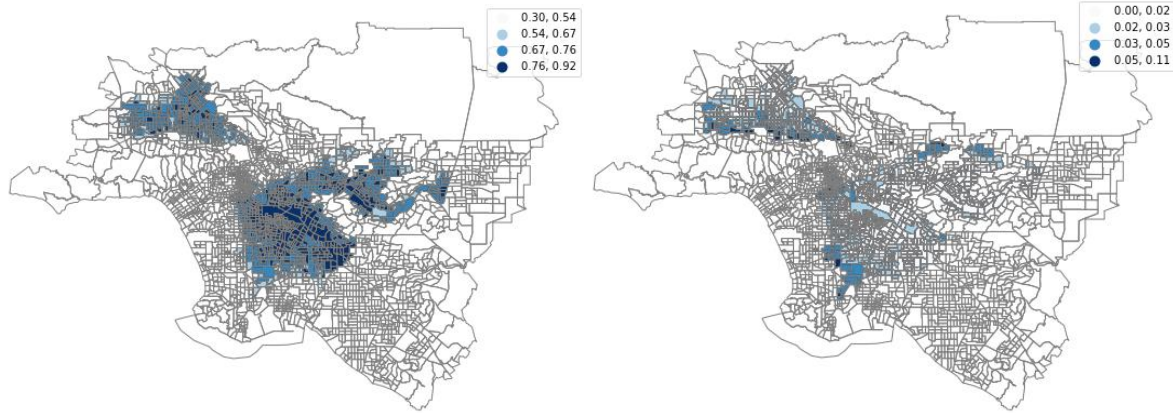

(C) The ratio of trips from high UH to high UH (D) The ratio of trips from high UH to low UH

**Figure 2.** Urban Heat Traps and Trips in Los Angeles Metropolitan Area. (A) shows that 52 percent of tracts are in high UH areas across Los Angeles. (C) 81 percent of tracts in high UH areas have trips to other high UH tracts, representing that Los Angeles is a metropolitan area with urban heat traps.

Similarly, the Chicago metropolitan area shows strong urban heat traps as well. Figure 3A maps the UH in Chicago. Chicago has 16 percent of its tract in low UH areas, while 49 percent of its tracts are in high UH areas. Figure 3B shows the ratio of trips visiting from low UH tracts to high UH tracts. A higher ratio of trips 0.17 to 0.24 occurs on the coast of Lake Michigan, meaning that a significant number of people living in low UH areas are visiting high UH areas on the coast of Lake Michigan. Figure 3D shows trips from high UH tracts to low UH tracts with a ratio as high as 0.08 to 0.13 occurring in the east. Figure 3C shows the ratio of trips visiting from high UH tracts to high UH tracts, with the ratio of trips from 0.44 to 0.91, which means that a large number of people living in high UH areas are visiting other high UH areas within the Chicago metropolitan area. About 78 percent of Chicago tracts in high UH areas have trips trapped inside high UH areas. Most of the UH traps are in the west of Chicago. At the same time, central Chicago presents an exceptionally high heat trap ratio, ranging from 0.79 to 0.91. Figure 3D. shows the ratio of trips visiting from high UH tracts to low UH tracts, with the ratio of trips from 0 to 0.13. This means that a relatively low number of people living in high UH areas are visiting low UH areas within the Chicago metropolitan area.

Comparing the UH traps between Chicago and Los Angeles, we can see that the traps in Chicago are clustered in one place, while in Los Angeles are distributed into multiple clusters.

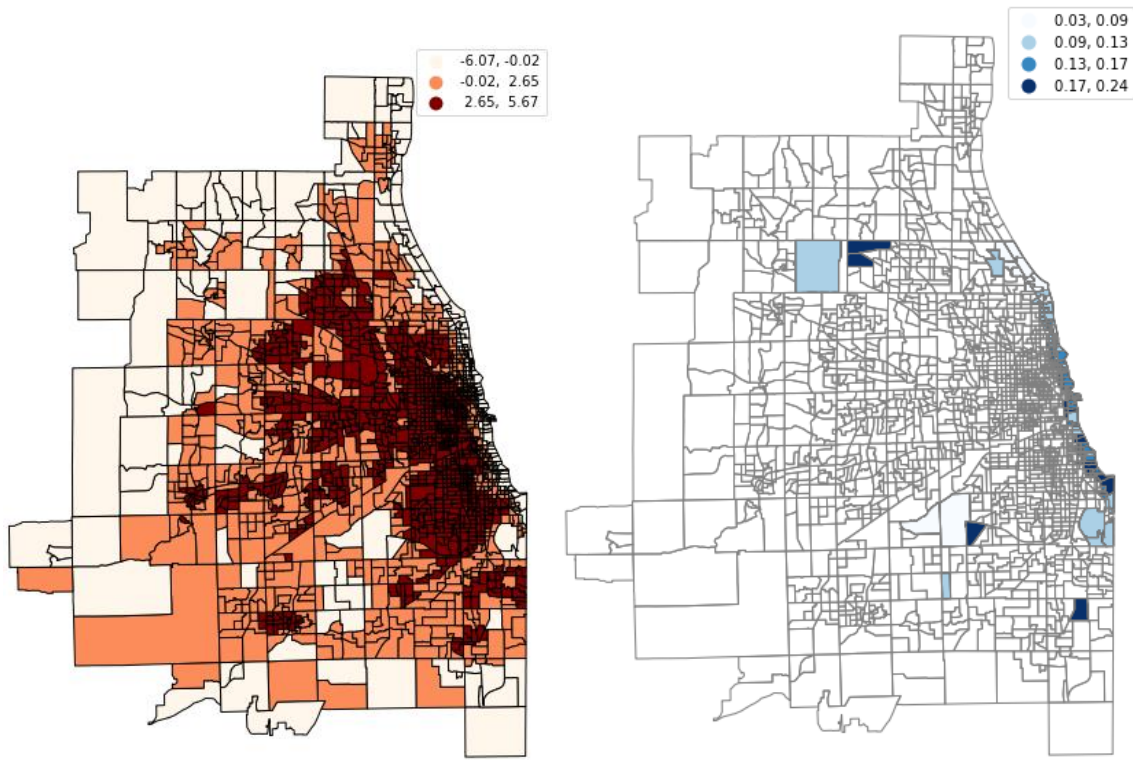

(A) Distribution of urban heat            (B) The ratio of trips from low UHI to high UHI

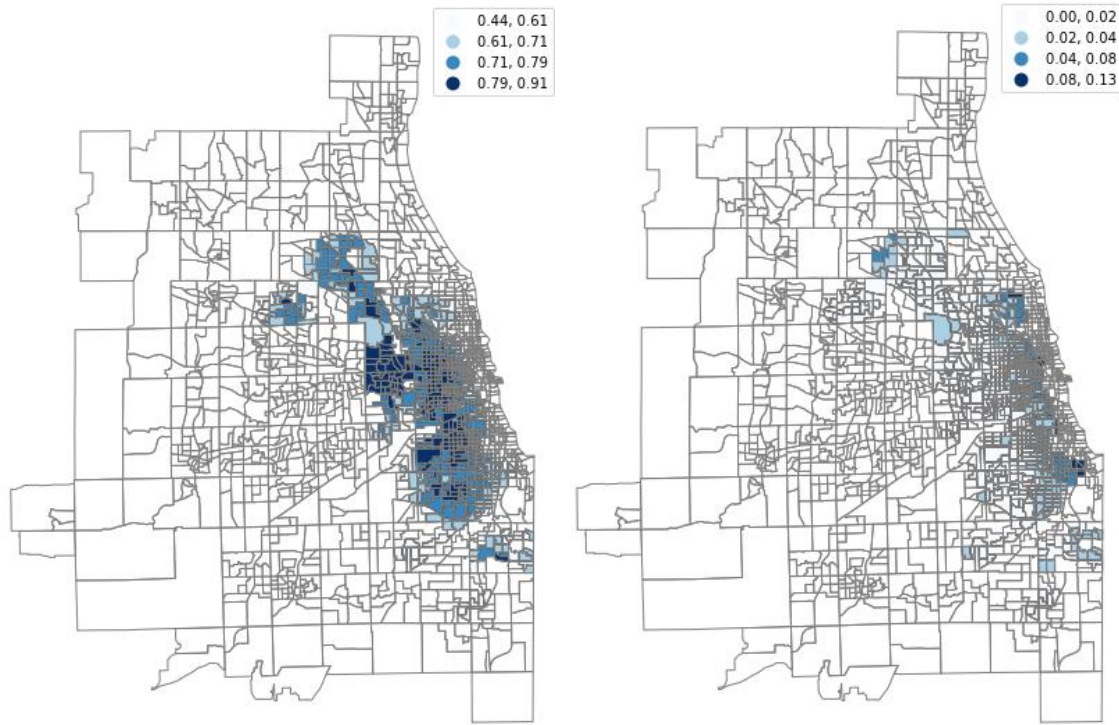

(C) The ratio of trips from high UHI to high UHI (D) The ratio of trips from high UHI to low UHI

**Figure 3**. UH Traps and Trips in the Chicago Metropolitan Area. (A) 16 percent of tracts are in low UH areas and 49 percent of tracts are in high UH areas across Chicago. (C) 78 percent of tracts in high UH areas have trips to high UH tracts, representing that Chicago is a metropolitan area with urban heat traps.

Figures 3 and 4 show that the Los Angeles and Chicago metropolitan areas both have significant urban heat traps. In Los Angeles, 52 percent of all the tracts are in high UH areas, while in Chicago, 49 percent are in high UH areas. The figures also show that trips from low UH areas to high UH areas are more frequent in the north of both cities, while trips from high UH areas to low UH areas are more common in the northwest and southwest of Los Angeles, and the east of Chicago. Additionally, the figures show that both cities present high heat trap trips, with around 80 percent of tracts with heat trap trips. This indicates that people in the high UH areas are likely not visiting the low UH areas to escape the heat, but instead are staying in other high UH areas.

**Cities with Low Urban Heat Traps**

Boston Metropolitan shows low urban heat traps. Figure 4A maps the UH in Boston. About 28 percent of tracts in Boston are in low UH areas, while 36 percent of tracts have high UH. Most of these high UH tracts are clustered in central Boston. Figure 4B shows the ratio of trips visiting from low UH tracts to high UH tracts. The ratio of such trips is from 0.04 to 0.19 and only occur in 4 percent of all the tracts with low UH. Figure 4C shows the ratio of trips visiting from high UH tracts to high UH tracts. This ratio ranges from 0.30 to 0.90. About 37 percent of tracts with high UH have trips trapped inside high UH areas. This percentage is relatively small when comparing to Los Angeles (81 percent) and Chicago (78 percent). Figure 4D shows the trips from high UH areas to low UH areas with ratio from 0 to 0.02. These results indicate that people living in low UH areas in the Boston metropolitan area are not frequently visiting high UH areas, which could be an indication of a fewer heat traps.

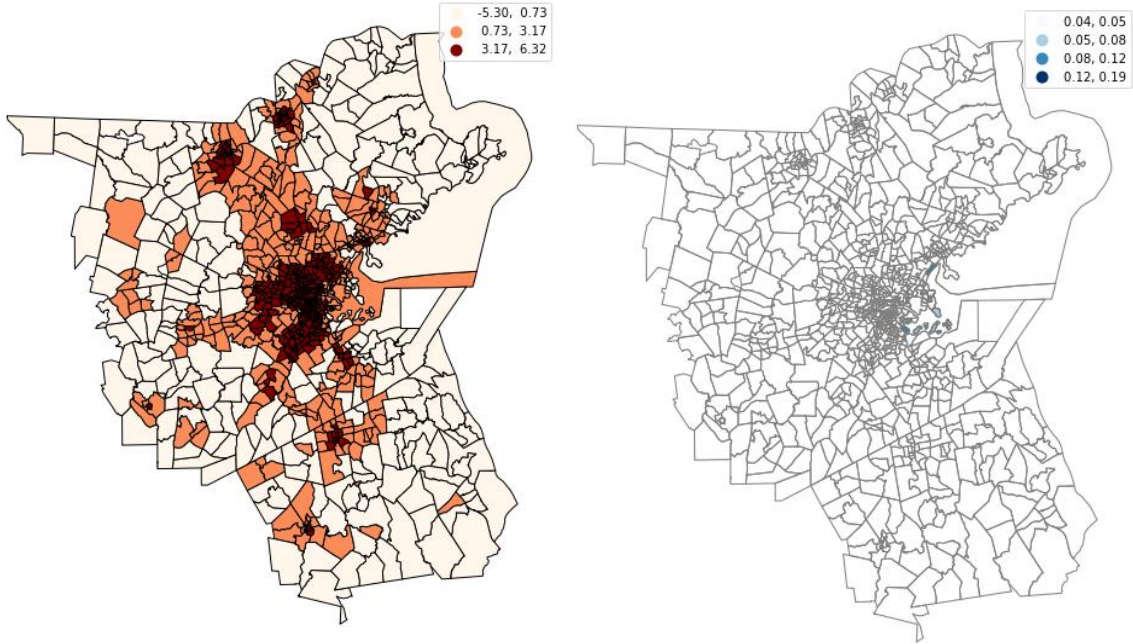

(A) Distribution of UH  (B) The ratio of trips from low UH to high UH

(C) The ratio of trips from high UH to high UH  (D) The ratio of trips from high UH to low UH

**Figure 4.** UH Traps and Trips in Boston Metropolitan Area. (A) 28 percent of tracts are low UH areas and 38 percent are in high UH areas across Boston. (C) 37 percent of tracts in high UH

areas have trips to high UH tracts, representing that Boston is a metropolitan area with low urban heat traps.

Similarly, the Atlanta Metropolitan also shows low UH traps. Figure 5A maps the UH in Atlanta. Atlanta has 32 percent of the tracts in low UH areas, while 21 percent are in high UH areas. Figure 5B shows the ratio of trips visiting from low UH tracts to high UH tracts. The ratio of such trips is from 0.01 to 0.16 and only occurred in 13 percent of all the low UH tracts. Figure 5C shows the ratio of trips visiting from high UH tracts to high UH tracts with ratios from 0.37 to 0.87. About 40 percent of high UH tracts have heat trap trips. This number is similar with Boston and is relatively small comparing to Los Angeles and Chicago. Figure 5D shows the trips from high UH areas to low UH areas, ranging from 0.01 to 0.12. This ratio is small but more significant than that of Boston, which means that comparing to Boston, more heat escape trips exist in Atlanta.

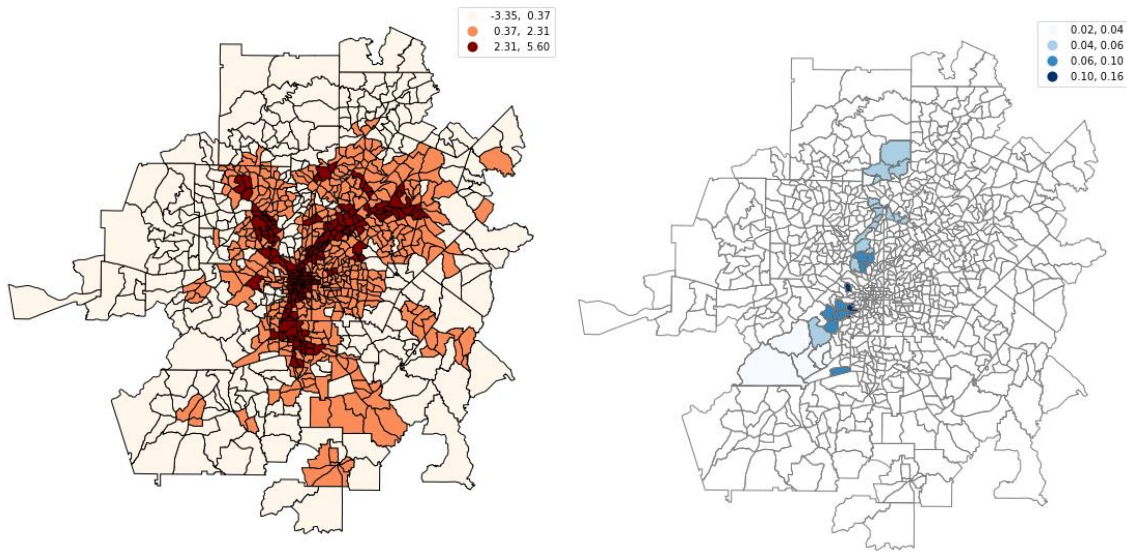

(A) Distribution of UH  (B) The ratio of trips from low UH to high UH

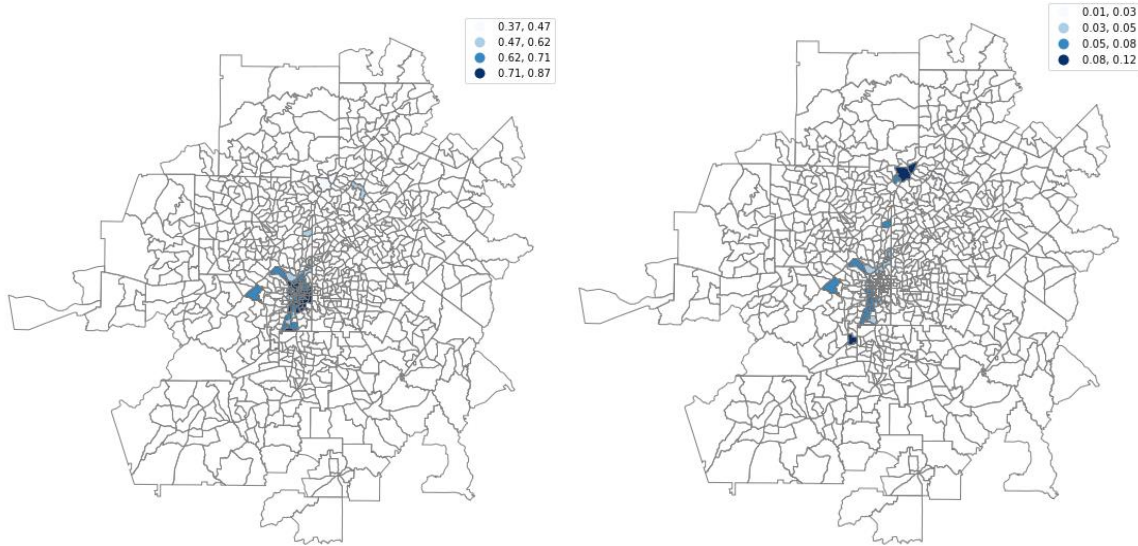

(C) The ratio of trips from high UH to high UH (D) The ratio of trips from high UH to low UH

**Figure 5**. UH Traps and Trips in Atlanta Metropolitan Area. (A) 32 percent of tracts are low UH and 21 percent are high UH across Atlanta. (C)40 percent of tracts in high UH areas have trips to high UH tracts, representing that Atlanta is a metropolitan area with low UH traps.

Figures 4 and 5 show that both Boston and Atlanta have relatively low UH comparing to Los Angeles and Chicago. In Boston, only 36 percent of tracts are in high UH, while in Atlanta, only 21 percent of the tracts are in high UH. The figures also show that trips from low UH areas to high UH areas are relatively rare in both cities, only 4 percent and 13 percent in low UH tracts in Boston and Atlanta, respectively. In both cities, the percentages of tracts with trips trapped inside high UH areas are lower than in Los Angeles and Chicago.

**Cities with High Urban Heat Escapes**

The Minneapolis Metropolitan Area shows high UH escapes. Figure 6A maps the UH in Minneapolis. The metropolitan area has 18 percent of its tracts in low UH areas, while 45 percent are in high UH areas. Figure 6B shows the ratio of trips from low UH tracts to high UH tracts. This ratio ranges from 0.03 to 0.34, occurring in 26 percent of low UH tracts. Figure 6C shows the ratio of trips from high UH tracts to high UH tracts with the ratios from 0.41 to 0.86. Figure 6D shows the ratio of trips from high UH tracts to low UH tracts. This ratio is between 0.01 and 0.13, occurring in 56 percent of high UH tracts. Comparing this high UH to low UH ratio with other cities, Minneapolis shows strong UH escape, indicating that a significant number of people living in high UH areas are visiting low UH areas in the Minneapolis metropolitan area.

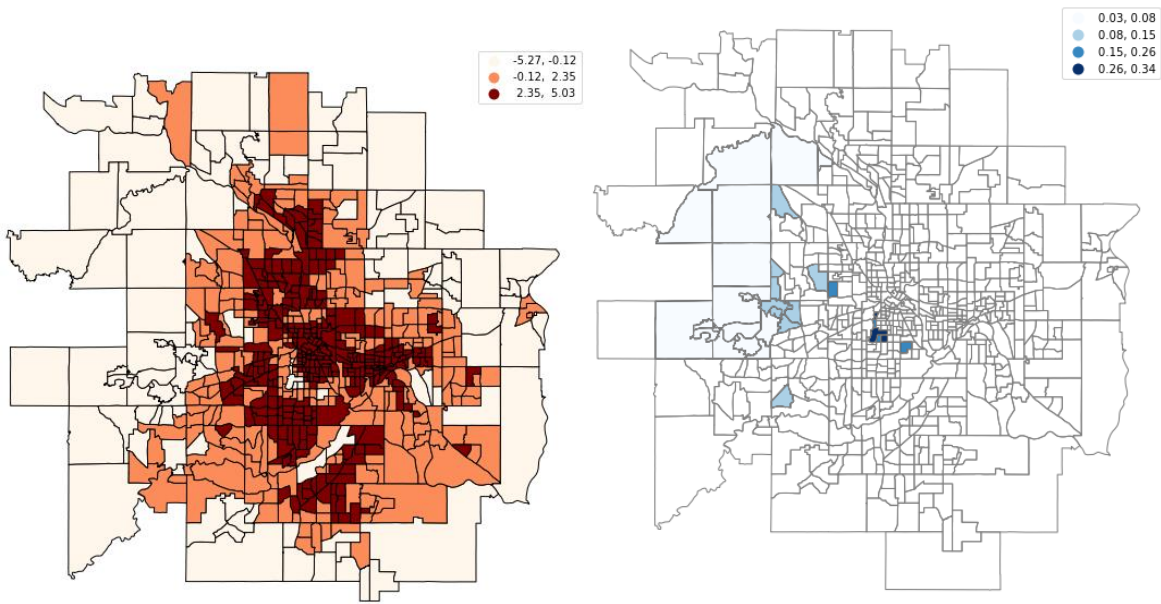

(A) Distribution of UH        (B) The ratio of trips from low UH to high UH

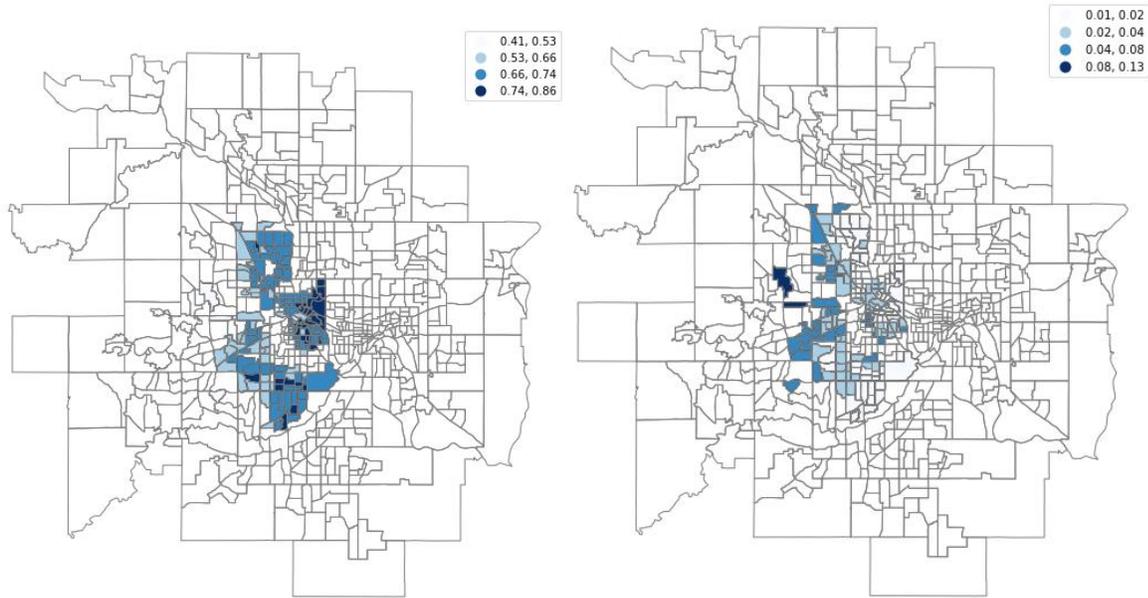

(C) The ratio of trips from high UH to high UH (D) The ratio of trips from high UH to low UH

**Figure 6.** UH Traps and Trips in Minneapolis Metropolitan Area. (A) 18 percent of tracts are low UH areas and 45 percent are in high UH areas across Minneapolis. (D) 56 percent of tracts in high UH areas have trips to low UH tracts, representing that Minneapolis has high heat escapes trips.

Similarly, the Dallas Metropolitan Area also shows high heat escapes. Figure 7A maps the UH in Dallas. Dallas has 10 percent of its tracts in low UH areas, while 50 percent of its tracts are in high UH areas. Figure 7A shows that the high UH tracts form multiple clusters across the city. Figure 7B shows the ratio of trips from low UH tracts to high UH tracts. This ratio is between 0.07 and 0.28, occurring in 38 percent of the low UH tracts. Figure 7C shows the ratio of trips from high UH tracts to high UH tracts with ratios from 0.39 to 0.82. Figure 7D shows the ratio of trips from high UH tracts to low UH tracts. The ratio of trips from high UH tracts to low UH tracts is notable, ranging from 0.00 to 0.16, in 49 percent of high UH tracts. This indicate that
Dallas has strong urban heat escapes trips.

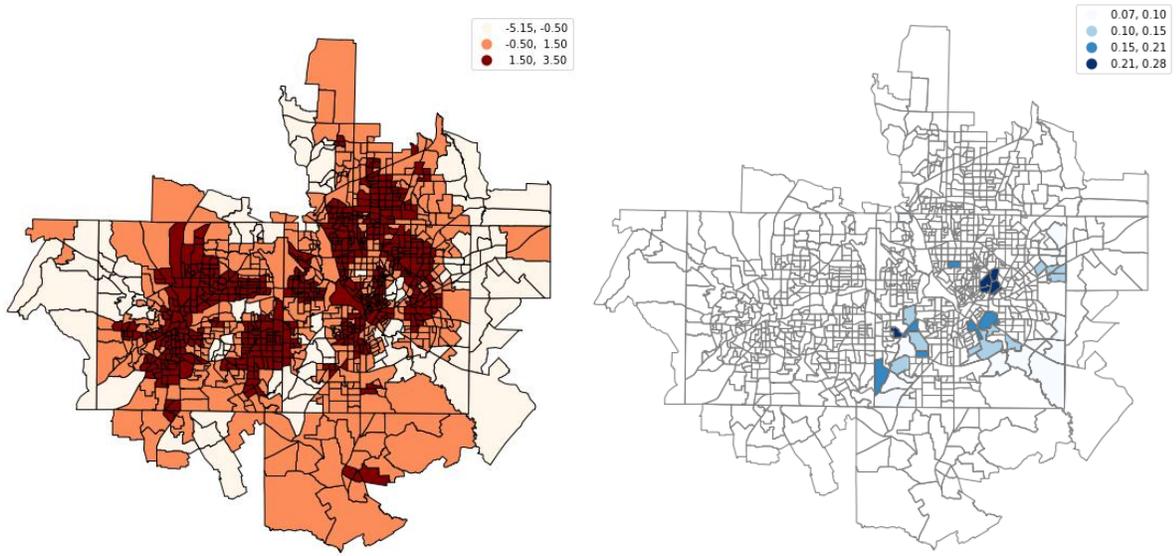

(A) Distribution of UH  (B) The ratio of trips from low UH to high UH

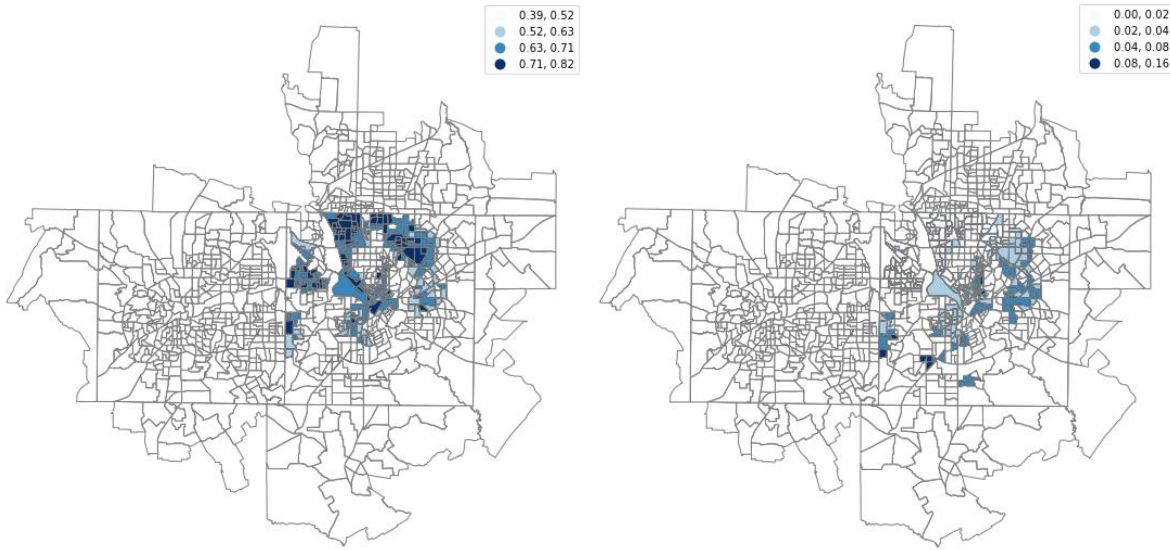

(C) The ratio of trips from high UH to high UH  (D) The ratio of trips from high UH to low UH

**Figure 7.** UH Traps and Trips in the Dallas Metropolitan Area. (A) 10 percent of tracts are low UH areas and 51 percent are in high UH areas across Dallas. (D) 49 percent of high UH tracts

have trips to low UH tracts, representing that Dallas is a metropolitan area with high urban heat escapes.

Figures 6 and 7 show that both Minneapolis and Dallas have significant urban heat escapes, with a higher ratio of trips from high UH tracts to low UH tracts when comparing to other metropolitan areas, such as Boston (37 percent) and Atlanta (24 percent). This indicates that people in the high UH areas travel to the low UH areas to escape the heat.

Additionally, this study offered important insights by examining the factors of distinctive characteristics underline spatial structures (Angel & Blei, 2016), facility distribution (Pereira et al., 2013), income, and racial segregation, as in Appendix C. However, no statistical significance was found between heat traps and attributes of demographic segregation. This interpolates that an urban heat trap is an emergent property (Georgiou, 2003) that cannot be attributed to the centrality of city facilities and demographics. Therefore, we observe that human mobility leads to the creation of traps, not escapes or escalates. Maybe people are more likely to go to places where they are more familiar.

**Discussion and Concluding Remarks**

This study utilized large-scale, high-resolution location-intelligence data to identify and quantify the urban heat (UH) exposure and people's response based on human mobility networks in urban areas. This study analyzed the intersection of UH and human mobility by examining the UH dataset and trips between tracts in February 2020 in twenty metropolitan areas. The study identified and analyzed three properties: heat traps, heat escapes, and heat escalate by quantifying the trips between tracts in high UH areas and low UH areas. This study found that not many cities have heat escapes or heat escalates trips. Heat escapes were found in Minneapolis and heat escalates were found in Los Angeles. A potential reason might be that people are more likely to stay in their resident areas.

Researchers and professionals are well aware of the diverse effects that UH can have heat-related diseases, such as respiratory difficulties among urban populations (Huang et al., 2020). However, there is little knowledge about the extent to which human mobility exacerbates UH. This study offers an innovative, data-driven method and metrics for using large-scale location intelligence data to assess UH exposure. This study evaluates the intersection of human mobility and the spatial distribution of urban heat. In addition, this study defines three important characteristics of people's potential response to UH based on trip destinations. Specifically, heat traps refer to population residing in high UH areas visit other high UH areas; heat escalates refer to population residing in low UH areas visit high UH areas and thus escalate their heat exposure; and heat escapes refer to population residing in high UH areas visit low UH areas and thus escape from their local heat. Defining these three different responses to UH can help researchers understand different characteristics of the urban areas.

There are several limitations of this study. First, this study is based on smartphone data. Smartphone users who allowed such location data collection is a biased sample. Visitors who do not own smartphones, such as children, teenagers, the elderly, and those with lower income, were less likely to be included in the data, which may create biases (Esmalian et al., 2021, Song et al., 2022). Additionally, efforts could be made to ensure that the sample of smartphone users is representative of the population as a whole, such as by using stratified sampling or weighting the data to account for any biases. We partially address this limitation by utilizing Spectus data, which has been demonstrated to contain a representative sample of users (Li & Mostafavi, 2022). Second, the mobility data does not include the visiting time for the destinations, which may cause mis-labeling of trip purposes. Future researchers could leverage other sources of data, such as surveys or observational studies, to further validate traveling information.

This study offers important insights to city designers and city planners. The three important characteristics of traps, escalates, and escapes are likely related to how heat exposure can affect people in different parts of a city. Better understandings of people's movements and associated heat exposure can provide city planner information for future city development. These characteristics may include factors such as the availability of shade and other forms of shelter, the accessibility of air conditioning and other cooling mechanisms, and the presence of social networks and support systems that can help people cope with heat waves and other extreme weather events. By understanding these characteristics, it may be possible to develop strategies and interventions that can help reduce the risks associated with heat exposure in urban environments.

**Data Availability**

All data were collected through a CCPA- and GDPR-compliant framework and utilized for research purposes. The data that support the findings of this study are available from Spectus, but restrictions apply to the availability of these data, which were used under license for the current study. The data can be accessed upon request submitted on spectus.ai. Other data we use in this study are all publicly available.

**Code Availability**

The code that supports the findings of this study is available from the corresponding author upon request.

**Declaration of Interests:** none

**Acknowledgement**

This material is based in part upon work supported by the National Science Foundation under Grant CMMI-1846069 (CAREER), Texas A&M University X-Grant 699, and the Microsoft Azure AI for Public Health grant. The authors also would like to acknowledge the data support from Spectus. Any opinions, findings, conclusions or recommendations expressed in this material are those of the authors and do not necessarily reflect the views of the National Science Foundation, Texas A&M University, Microsoft Azure, or Spectus.

Farahmand, H., Wang, W., Mostafavi, A., & Maron, M. (2022). Anomalous human activity fluctuations from digital trace data signal flood inundation status. *Environment and Planning B: Urban Analytics and City Science*, *49*(7), 1893–1911. https://doi.org/10.1177/23998083211069990

Georgiou, I. (2003). The idea of emergent property. *Journal of the Operational Research Society*, *54*(3), 239–247. https://doi.org/10.1057/palgrave.jors.2601520

Glencross, D. A., Ho, T.-R., Camiña, N., Hawrylowicz, C. M., & Pfeffer, P. E. (2020). Air pollution and its effects on the immune system. *Free Radical Biology and Medicine*, *151*, 56–68. https://doi.org/10.1016/j.freeradbiomed.2020.01.179

Hu, Y., Hou, M., Jia, G., Zhao, C., Zhen, X., & Xu, Y. (2019). Comparison of surface and canopy urban heat islands within megacities of eastern China. *ISPRS Journal of Photogrammetry and Remote Sensing*, *156*, 160–168. https://doi.org/10.1016/j.isprsjprs.2019.08.012

Huang, H., Yang, H., Deng, X., Zeng, P., Li, Y., Zhang, L., & Zhu, L. (2020). Influencing mechanisms of urban heat island on respiratory diseases. *Iranian Journal of Public Health*. https://doi.org/10.18502/ijph.v48i9.3023

Huang, X., Li, Z., Jiang, Y., Li, X., & Porter, D. (2020). Twitter reveals human mobility dynamics during the COVID-19 pandemic. *PLOS ONE*, *15*(11), e0241957. https://doi.org/10.1371/journal.pone.0241957

Hunter, R. F., Cleland, C., Cleary, A., Droomers, M., Wheeler, B. W., Sinnett, D., Nieuwenhuijsen, M. J., & Braubach, M. (2019). Environmental, health, wellbeing, social and equity effects of urban green space interventions: A meta-narrative evidence synthesis. *Environment International*, *130*, 104923. https://doi.org/10.1016/j.envint.2019.104923
31

**Appendices**

**Appendix A. Urban Heat and Human Mobility Ratios in Cities**

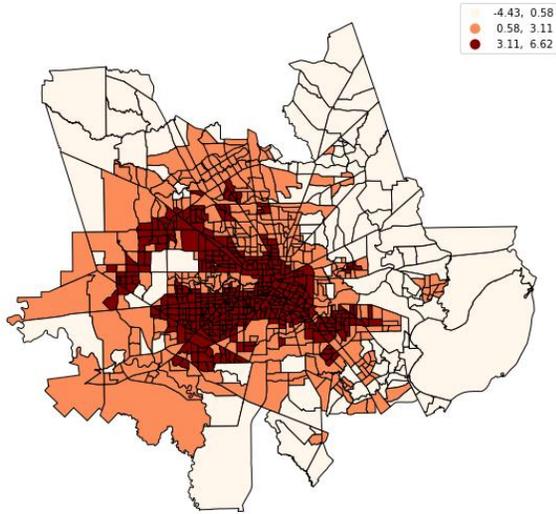
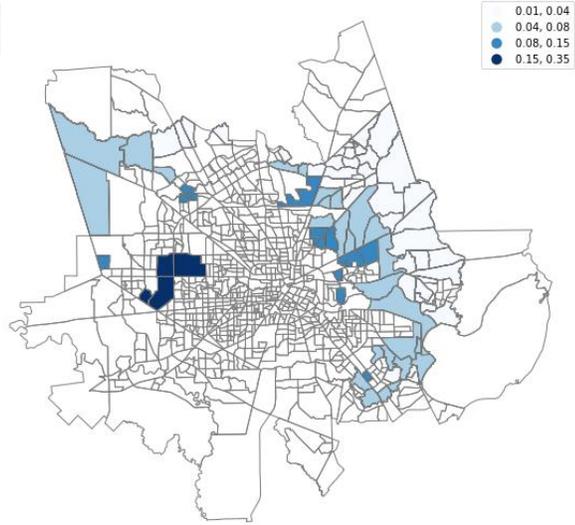

(A) Distribution of urban heat                (B) The ratio of trips from low UHI to high UHI

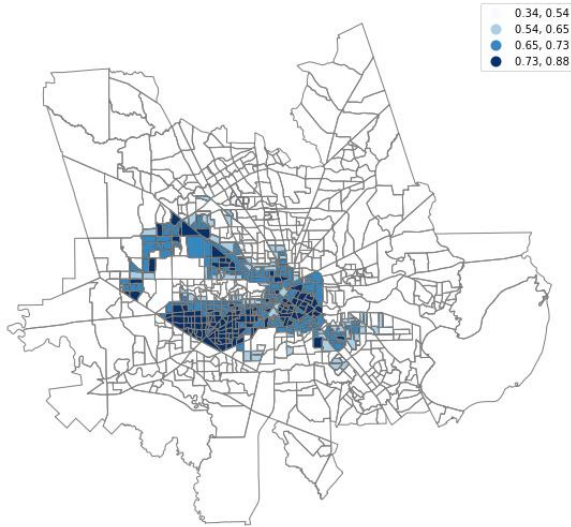
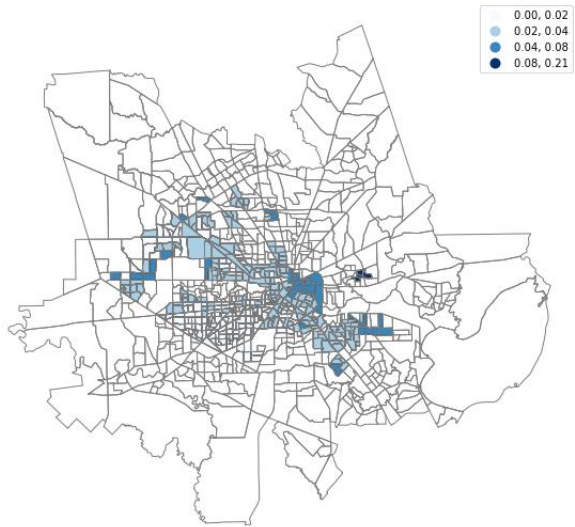

(C) The ratio of trips from high UHI to high UHI (D) The ratio of trips from high UHI to low UHI

**Figure A - 1**. Urban Heat Traps and Trips in Houston Metropolitan Area. (A) shows that 14 percent of census tracts are low UHI areas, and 48 percent are in high UHI areas across Houston. (C) 93 percent of census tracts in high urban heat areas have trips to high urban heat census tract, representing that Houston is a metropolitan area with high urban heat escapes.

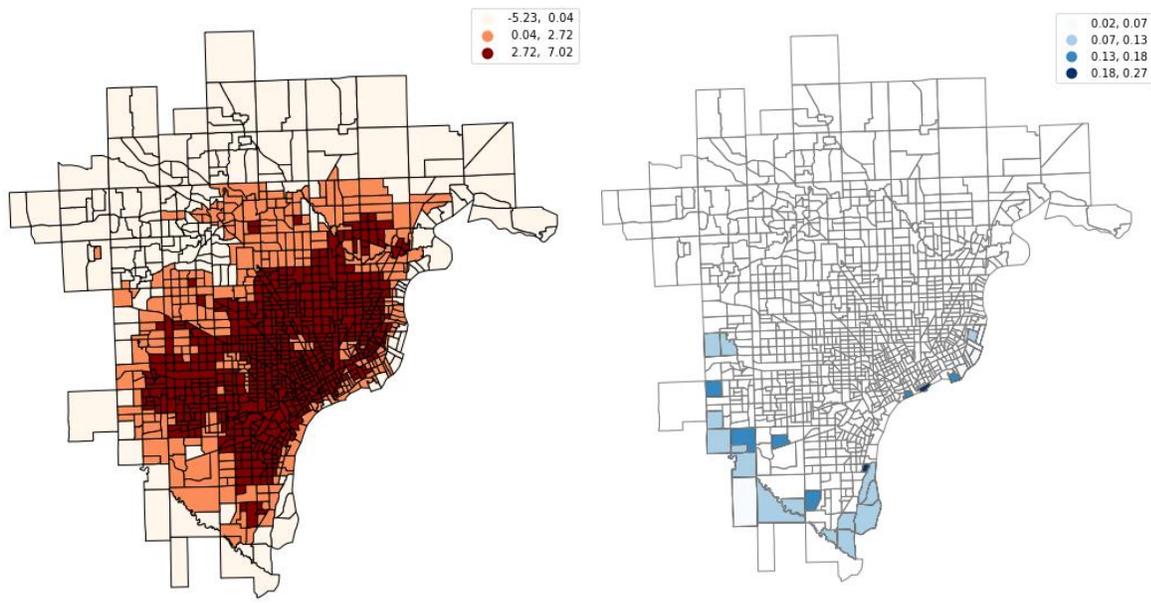

(A) Distribution of urban heat  (B) The ratio of trips from low UHI to high UHI

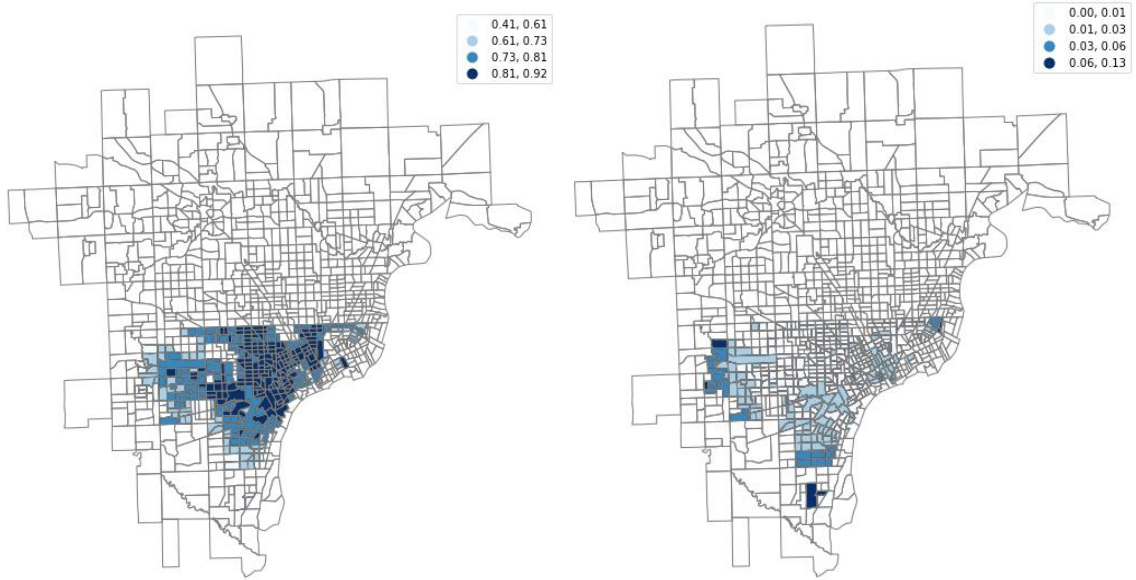

(C) The ratio of trips from high UHI to high UHI (D) The ratio of trips from high UHI to low UHI

**Figure A - 2.** Urban Heat Traps and Trips in Detroit Metropolitan Area. (A) shows that 16 percent of census tracts are low UHI areas, and 57 percent are in high UHI areas across Detroit.

(C) 62 percent of census tracts in high urban heat areas have trips to high urban heat census tract, representing that Detroit is a metropolitan area with high urban heat escapes.

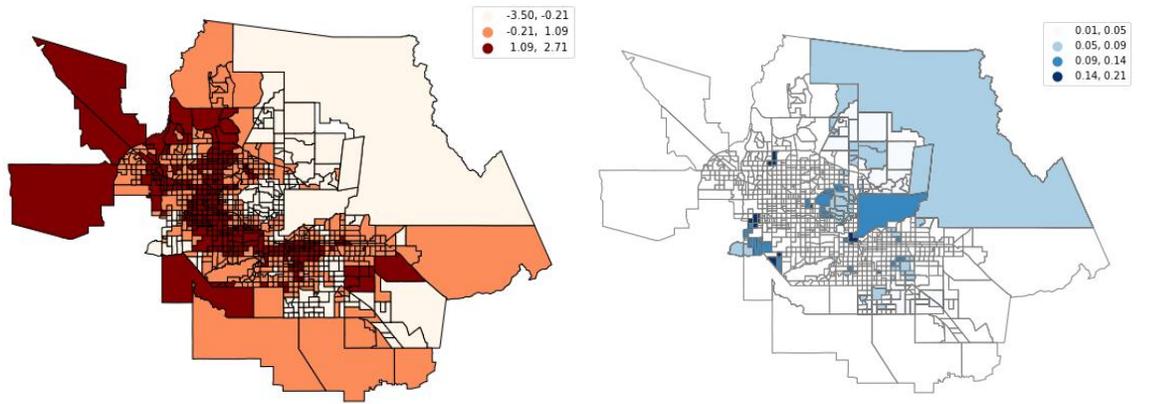

(A) Distribution of urban heat  (B) The ratio of trips from low UHI to high UHI

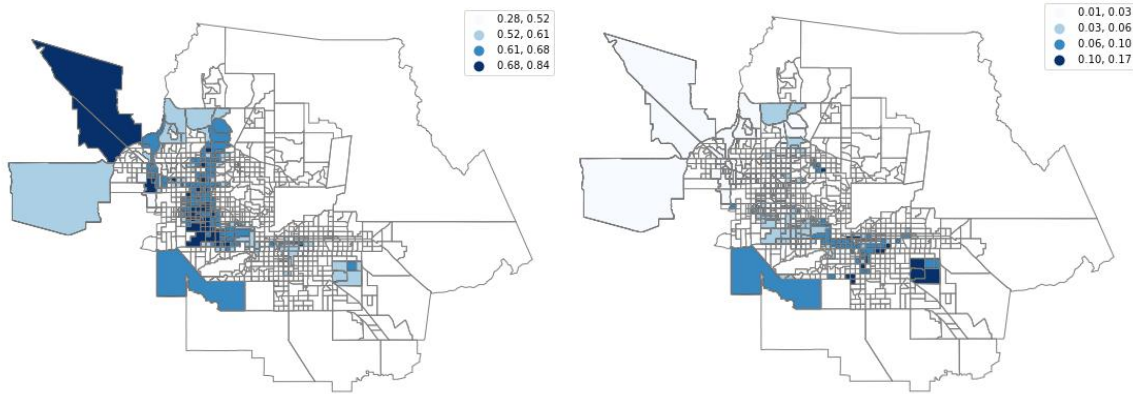

(C) The ratio of trips from high UHI to high UHI (D) The ratio of trips from high UHI to low UHI

**Figure A - 3.** Urban Heat Traps and Trips in Phoenix Metropolitan Area. (A) shows that 18 percent of census tracts are low UHI areas, and 37 percent are in high UHI areas across Phoenix. (B) 95 percent of census tracts in low urban heat areas have trips to high urban heat census tract with ratio of trips as high as 0.21. (C) 100 percent of census tracts in low urban heat areas have trips to high urban heat census tract, representing that Phoenix is a metropolitan area with high urban heat escapes and high urban heat escalates.

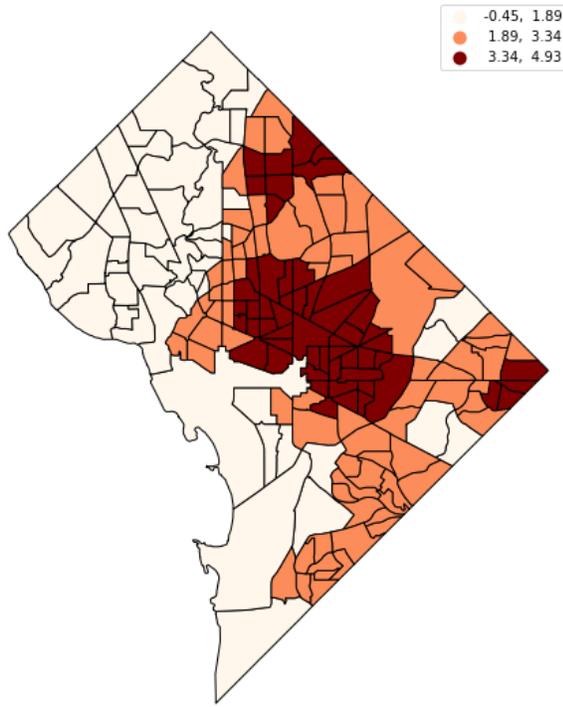
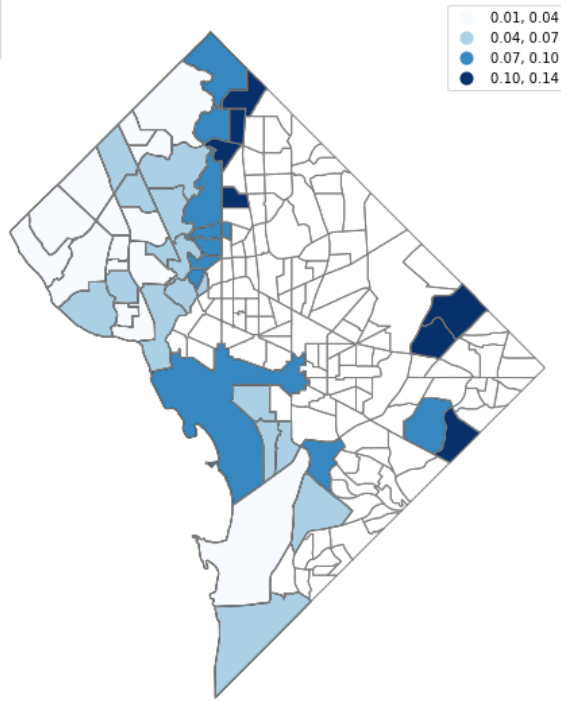

(A) Distribution of urban heat  (B) The ratio of trips from low UHI to high UHI

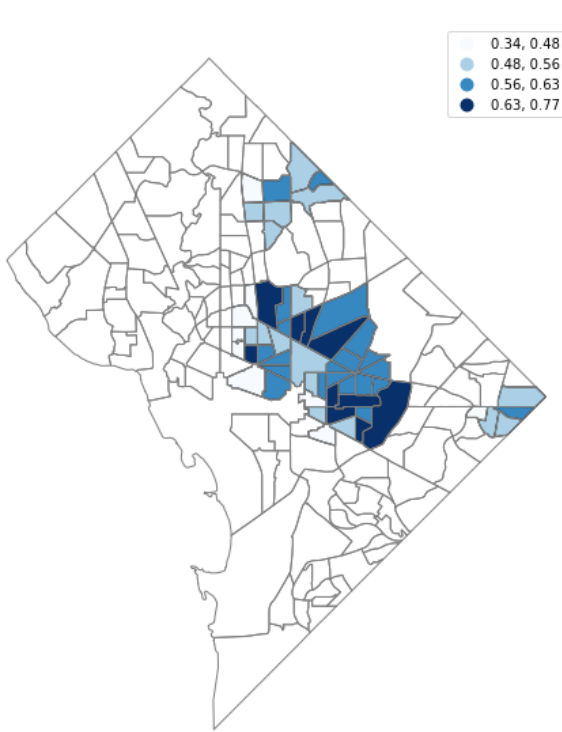
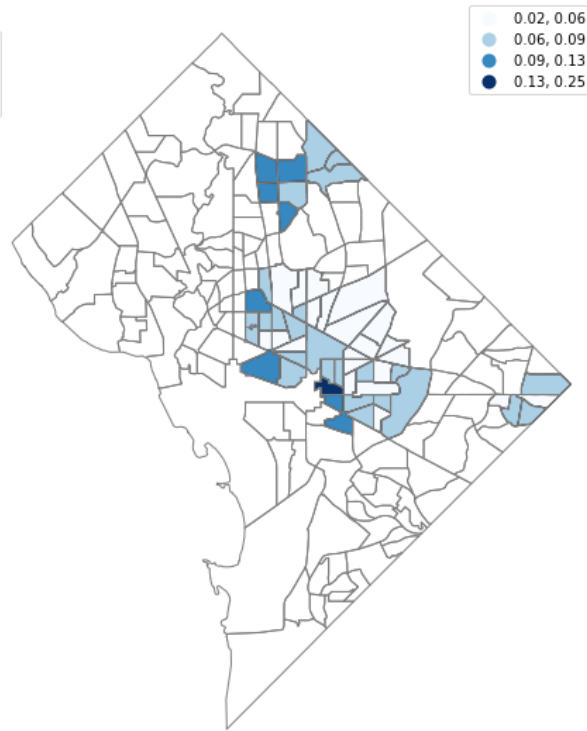

(C) The ratio of trips from high UHI to high UHI  (D) The ratio of trips from high UHI to low UHI

**Figure A - 4.** Urban Heat Traps and Trips in Washington DC Metropolitan Area. (A) shows that 27 percent of census tracts are low UHI areas, and 31 percent are in high UHI areas across Washington DC. (B) 100 percent of census tracts in low urban heat areas have trips to high urban heat census tract with ratio of trips as high as 0.14. (C) 100 percent of census tracts in low urban heat areas have trips to high urban heat census tract, representing that Washington DC is a metropolitan area with high urban heat escapes and high urban heat escalates.

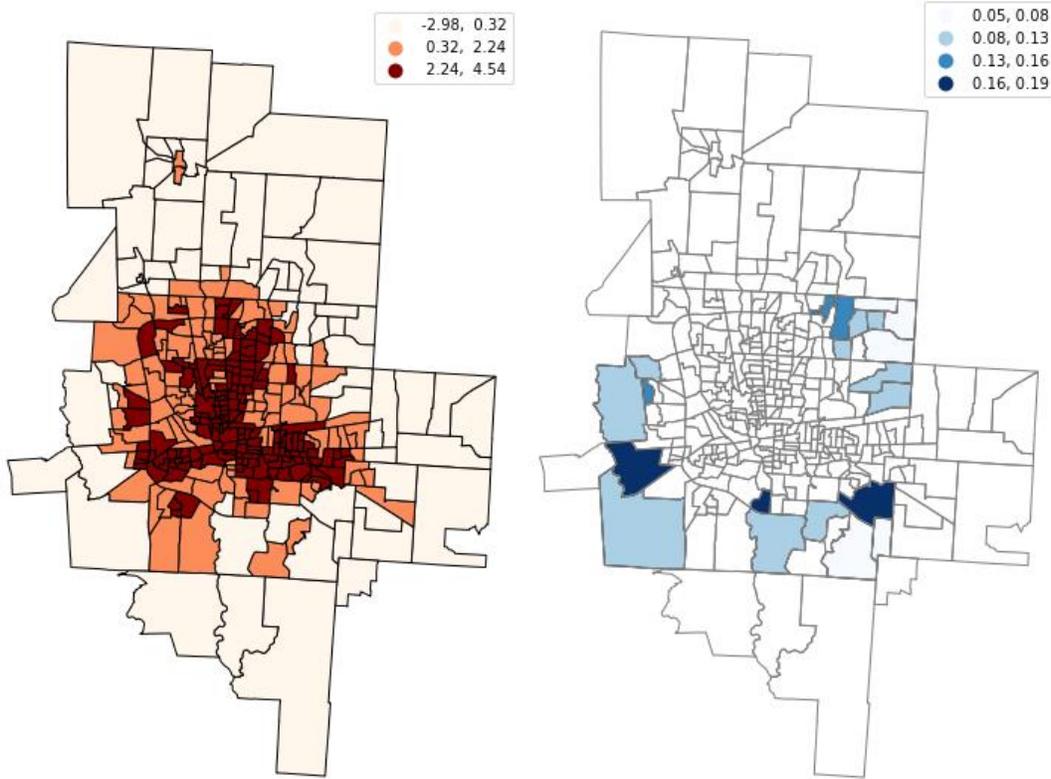

(A) Distribution of urban heat         (B) The ratio of trips from low UHI to high UHI

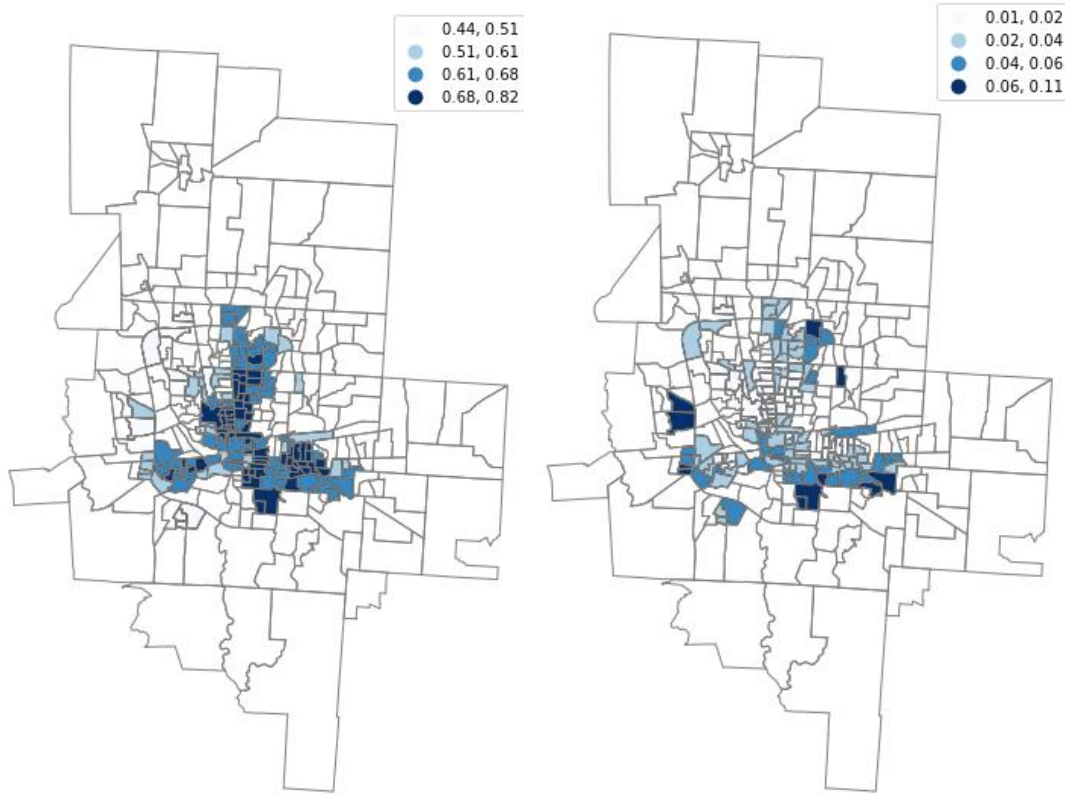

(C) The ratio of trips from high UHI to high UHI (D) The ratio of trips from high UHI to low UHI

**Figure A - 5.** Urban Heat Traps and Trips in Columbus Metropolitan Area. (A) shows that 20 percent of census tracts are low UHI areas, and 46 percent are in high UHI areas across Columbus. (B) 31 percent of census tracts in low urban heat areas have trips to high urban heat census tract with ratio of trips as high as 0.19. (C) 100 percent of census tracts in low urban heat areas have trips to high urban heat census tract, representing that Columbus is a metropolitan area with high urban heat escapes and high urban heat escalates.

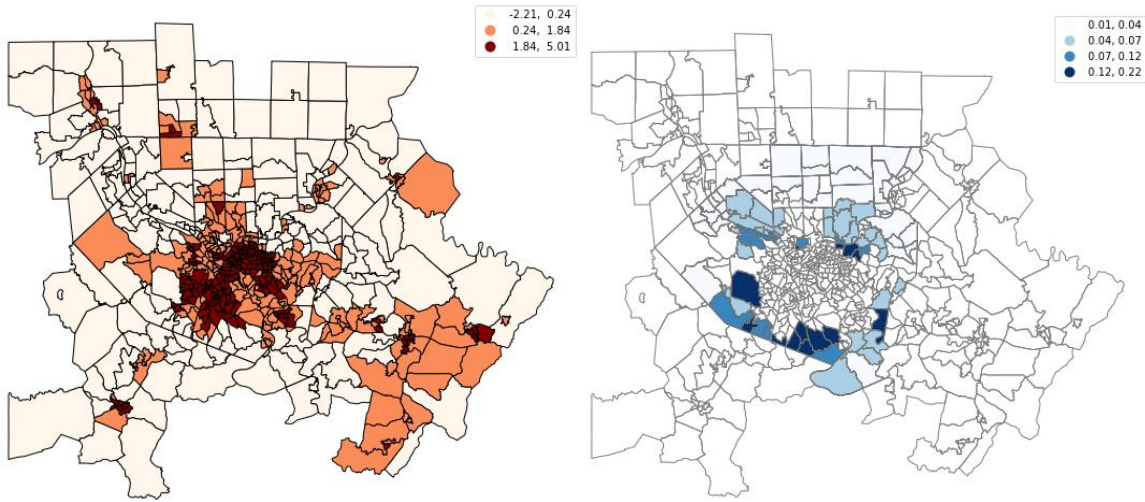

(A) Distribution of urban heat            (B) The ratio of trips from low UHI to high UHI

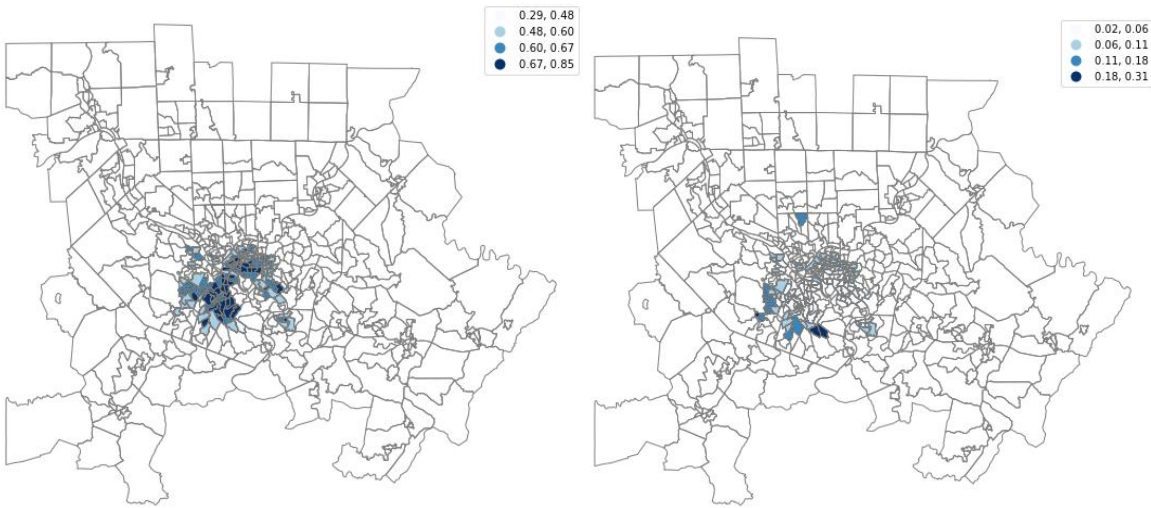

(C) The ratio of trips from high UHI to high UHI (D) The ratio of trips from high UHI to low UHI

**Figure A - 6.** Urban Heat Traps and Trips in Pittsburgh Metropolitan Area. (A) shows that 34 percent of census tracts are low UHI areas, and 32 percent are in high UHI areas across Pittsburgh. (B) 49 percent of census tracts in low urban heat areas have trips to high urban heat census tract with ratio of trips as high as 0.22. (C) 88 percent of census tracts in low urban heat areas have trips to high urban heat census tract, representing that Pittsburgh is a metropolitan area with high urban heat escalates and high urban heat traps.

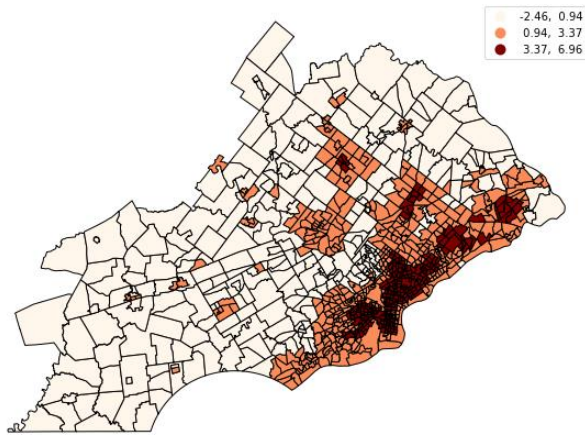 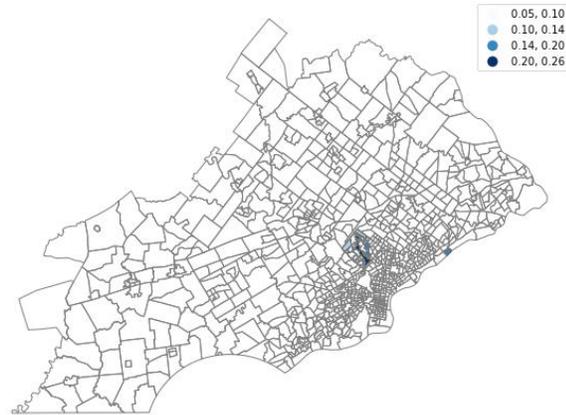

(A) Distribution of urban heat            (B) The ratio of trips from low UHI to high UHI

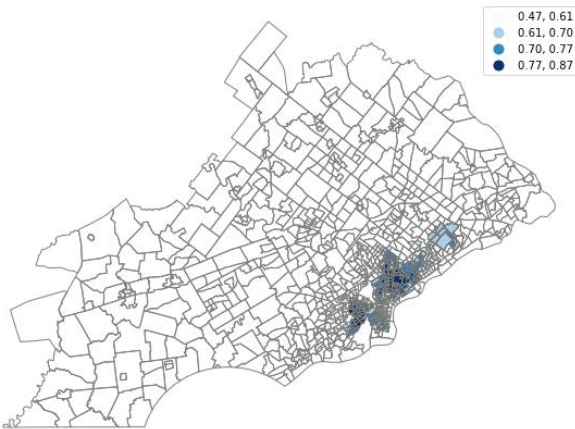 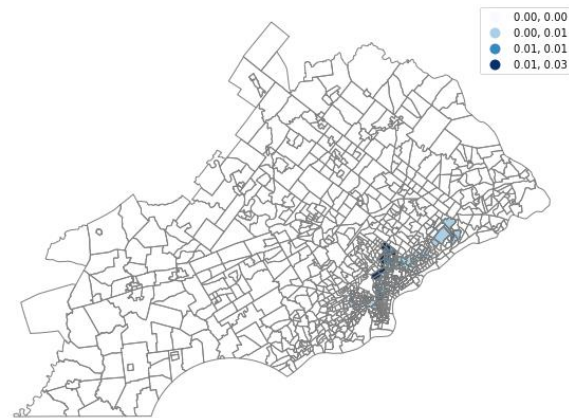

(C) The ratio of trips from high UHI to high UHI  (D) The ratio of trips from high UHI to low UHI

**Figure A - 7**. Urban Heat Traps and Trips in Philadelphia Metropolitan Area. (A) shows that 28 percent of census tracts are low UHI areas, and 29 percent are in high UHI areas across Philadelphia. (C) 85 percent of census tracts in low urban heat areas have trips to high urban heat census tract, representing that Philadelphia is a metropolitan area with high urban heat traps.

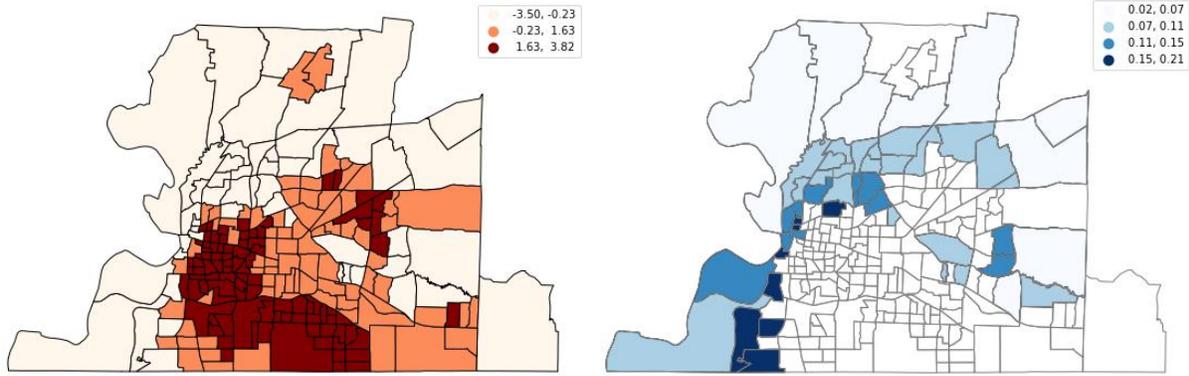

(A) Distribution of urban heat  (B) The ratio of trips from low UHI to high UHI

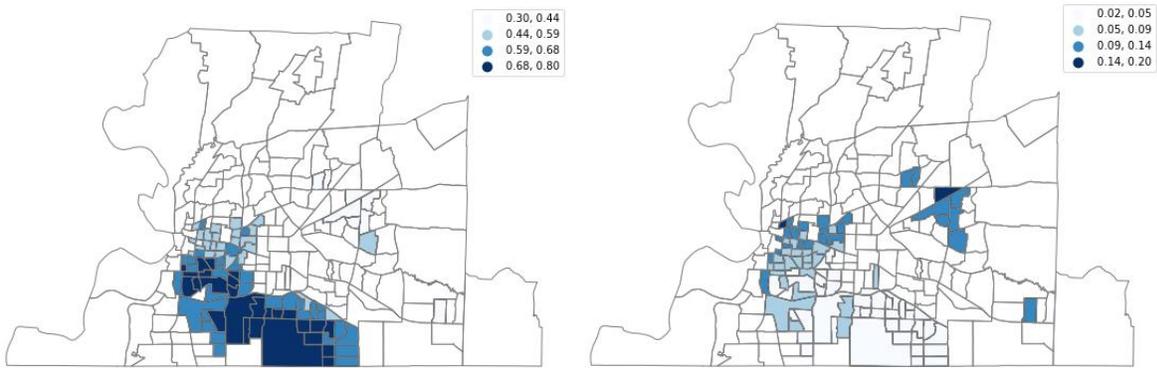

(C) The ratio of trips from high UHI to high UHI (D) The ratio of trips from high UHI to low UHI

**Figure A - 8**. Urban Heat Traps and Trips in Memphis Metropolitan Area. (A) shows that 23 percent of census tracts are low UHI areas, and 42 percent are in high UHI areas across Memphis. (B) 98 percent of census tracts in low urban heat areas have trips to high urban heat census tract with ratio of trips as high as 0.21 (C)100 percent of census tracts in low urban heat areas have trips to high urban heat census tract, representing that Memphis is a metropolitan area with high urban heat escalates and high urban heat traps.

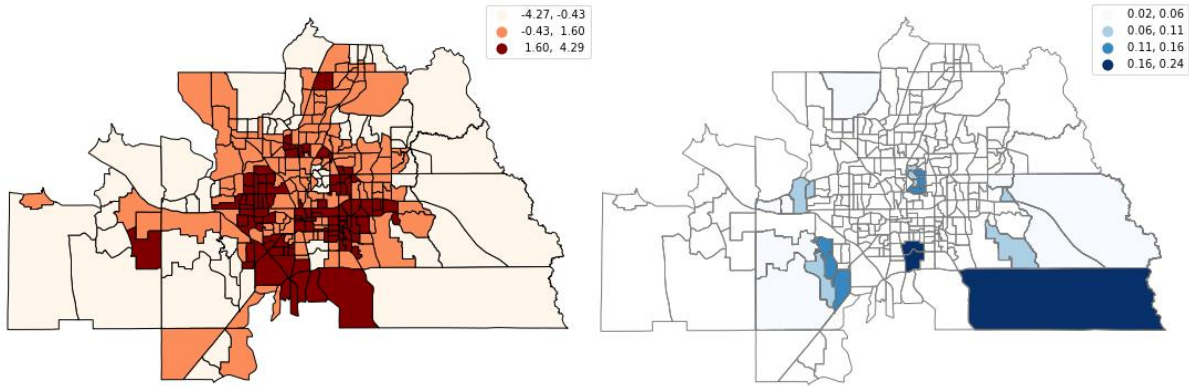

(A) Distribution of urban heat     (B) The ratio of trips from low UHI to high UHI

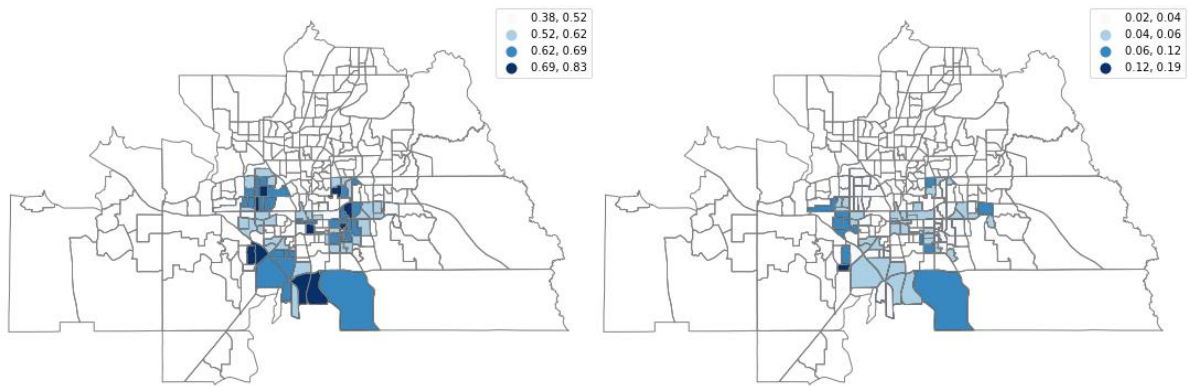

(C) The ratio of trips from high UHI to high UHI (D) The ratio of trips from high UHI to low UHI

**Figure A - 9**. Urban Heat Traps and Trips in Orlando Area. (A) shows that 19 percent of census tracts are low UHI areas, and 33 percent are in high UHI areas across Orlando. (B) 45 percent of census tracts in low urban heat areas have trips to high urban heat census tract with ratio of trips as high as 0.24 (C)89 percent of census tracts in low urban heat areas have trips to high urban heat census tract, representing that Orlando is a metropolitan area with high urban heat escalates and high urban heat traps.

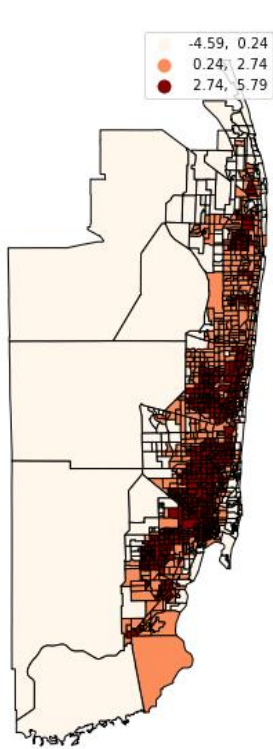 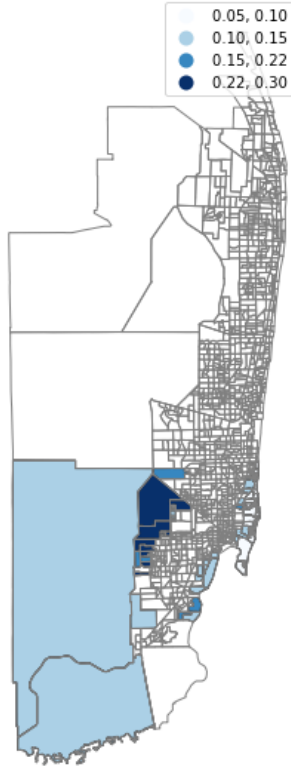

(A) Distribution of urban heat  (B) The ratio of trips from low UHI to high UHI

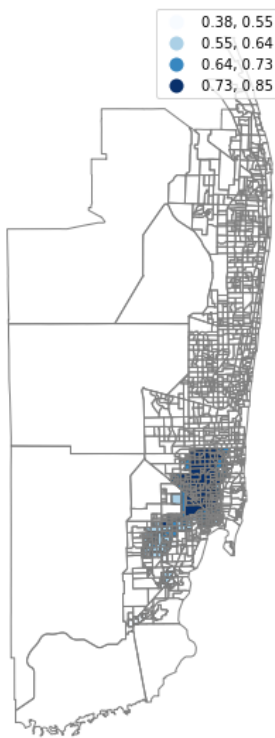 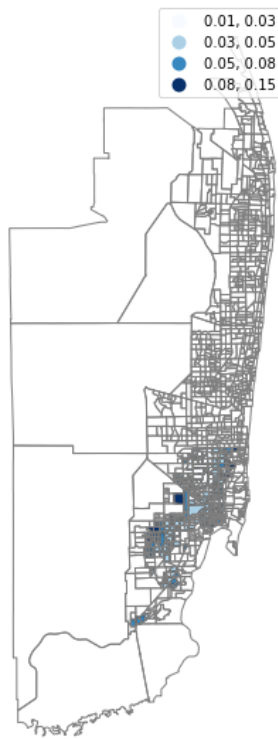

(C) The ratio of trips from high UHI to high UHI (D) The ratio of trips from high UHI to low UHI

**Figure A - 10**. Urban Heat Traps and Trips in Miami Area(A) shows that 23 percent of census tracts are low UHI areas, and 43 percent are in high UHI areas across Miami. (B) 42 percent of census tracts in low urban heat areas have trips to high urban heat census tract with ratio of trips as high as 0.30 (C)57 percent of census tracts in low urban heat areas have trips to high urban heat census tract, representing that Miami is a metropolitan area with high urban heat escalates and high urban heat traps.

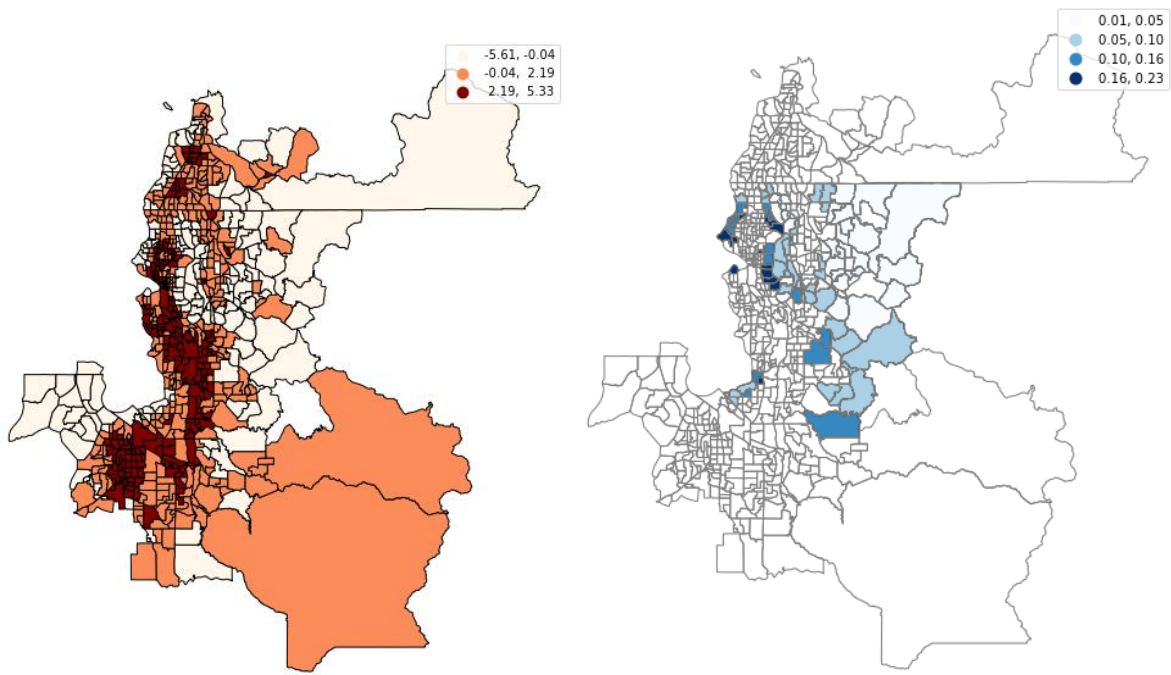

(A) Distribution of urban heat        (B) The ratio of trips from low UHI to high UHI

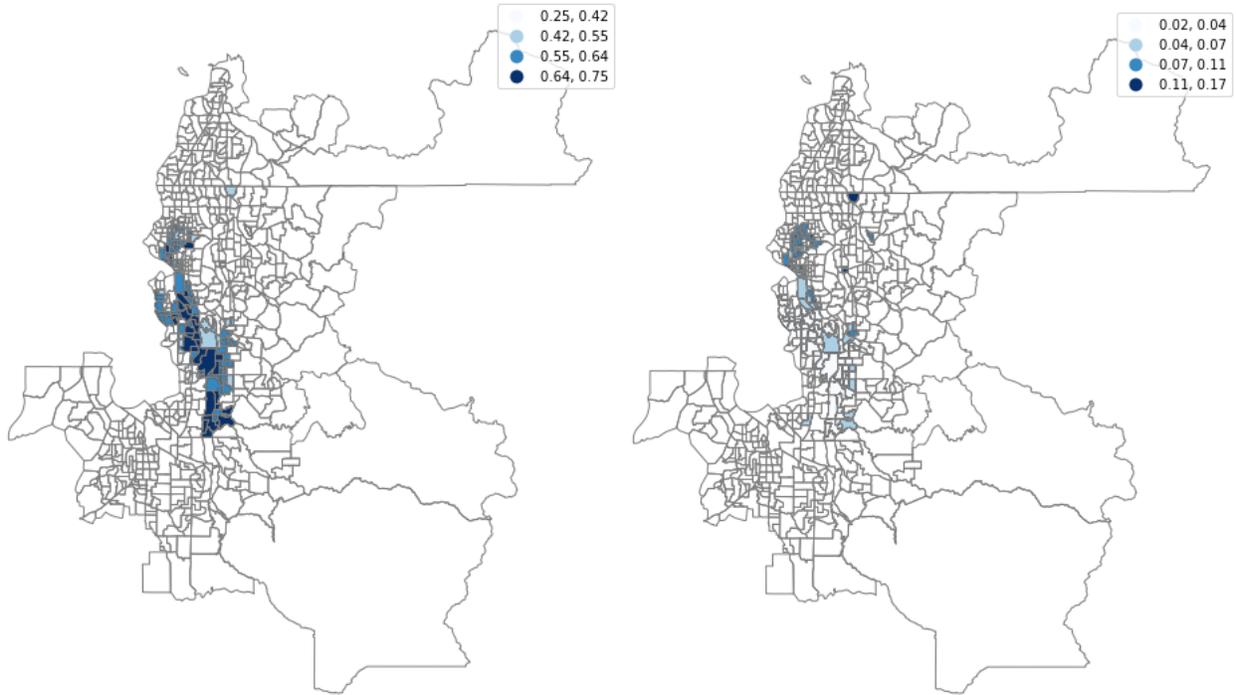

(C) The ratio of trips from high UHI to high UHI (D) The ratio of trips from high UHI to low UHI

**Figure A - 11.** Urban Heat Traps and Trips in Seattle Area. (A) shows that 23 percent of census tracts are low UHI areas, and 30 percent are in high UHI areas across Seattle. (C)60 percent of census tracts in low urban heat areas have trips to high urban heat census tract, representing that Seattle is a metropolitan area with low urban heat traps.

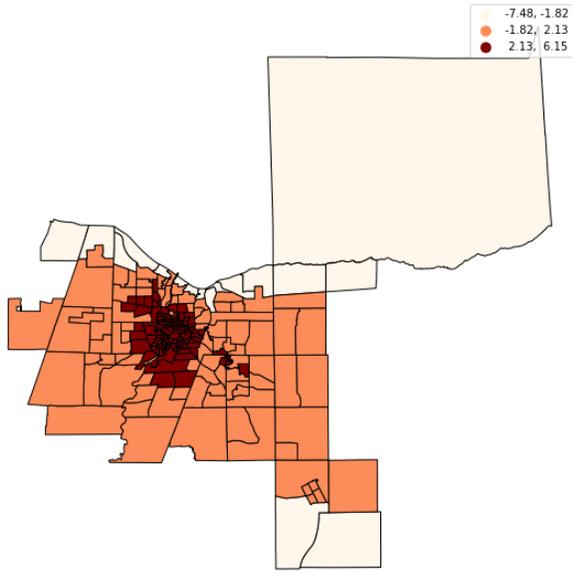
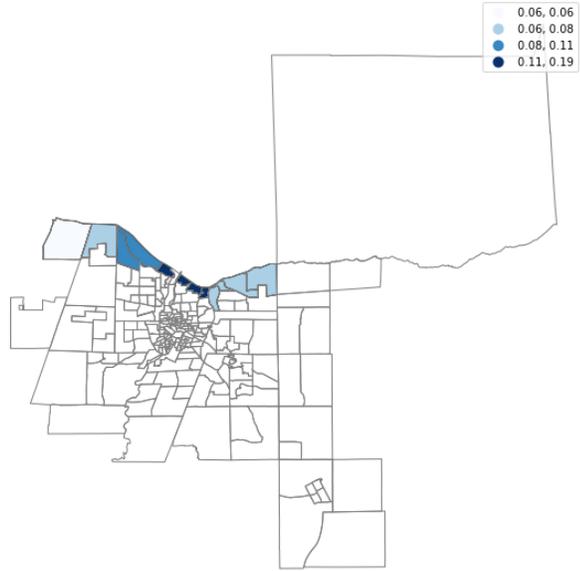

(A) Distribution of urban heat    (B) The ratio of trips from low UHI to high UHI

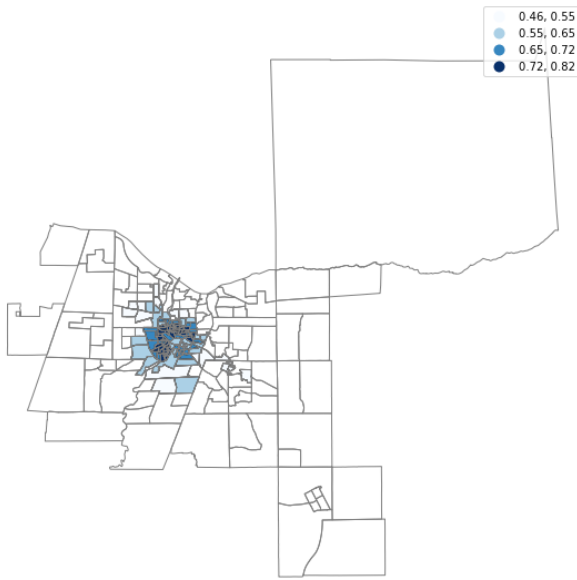
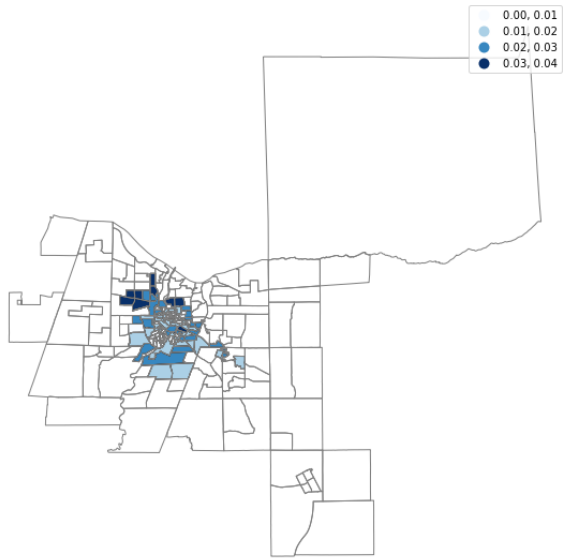

(C) The ratio of trips from high UHI to high UHI  (D) The ratio of trips from high UHI to low UHI

**Figure A - 12.** Urban Heat Traps and Trips in Rochester Area. (A) shows that 8 percent of census tracts are low UHI areas, and 50 percent are in high UHI areas across Rochester. (B) 71 percent of census tracts in low urban heat areas have trips to high urban heat census tract with ratio of trips as high as 0.19 (C) 100 percent of census tracts in low urban heat areas have trips to high urban heat census tract, representing that Rochester is a metropolitan area with high urban heat escalates and high urban heat traps.

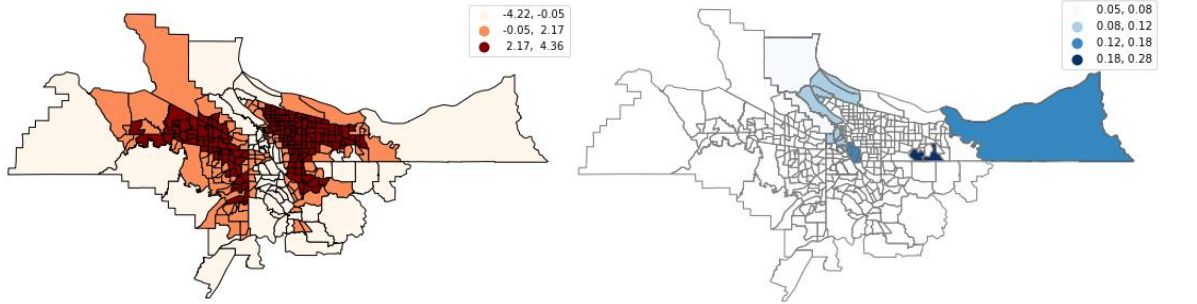

(A) Distribution of urban heat  (B) The ratio of trips from low UHI to high UHI

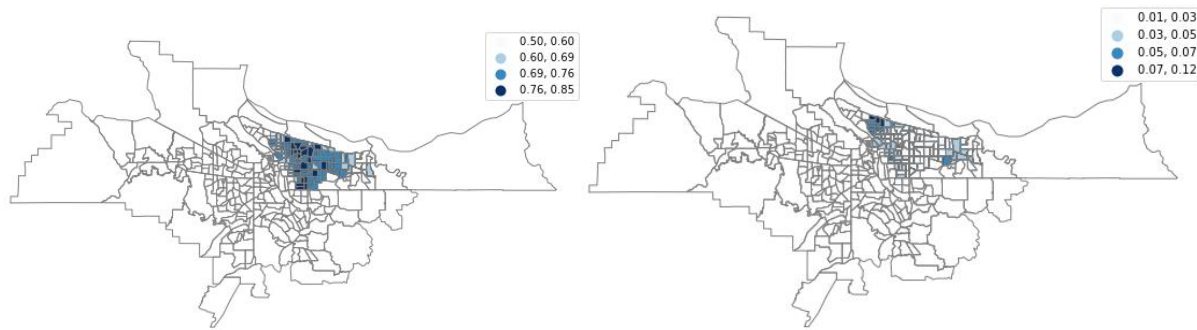

(C) The ratio of trips from high UHI to high UHI (D) The ratio of trips from high UHI to low UHI

**Figure A - 13**. Urban Heat Traps and Trips in Portland Area. (A) shows that 17 percent of census tracts are low UHI areas, and 53 percent are in high UHI areas across Portland. (B) 43 percent of census tracts in low urban heat areas have trips to high urban heat census tract with ratio of trips as high as 0.28. (C)60 percent of census tracts in high urban heat areas have trips to high urban heat census tract, representing that Portland is a metropolitan area with high urban heat escalates and high urban heat traps.

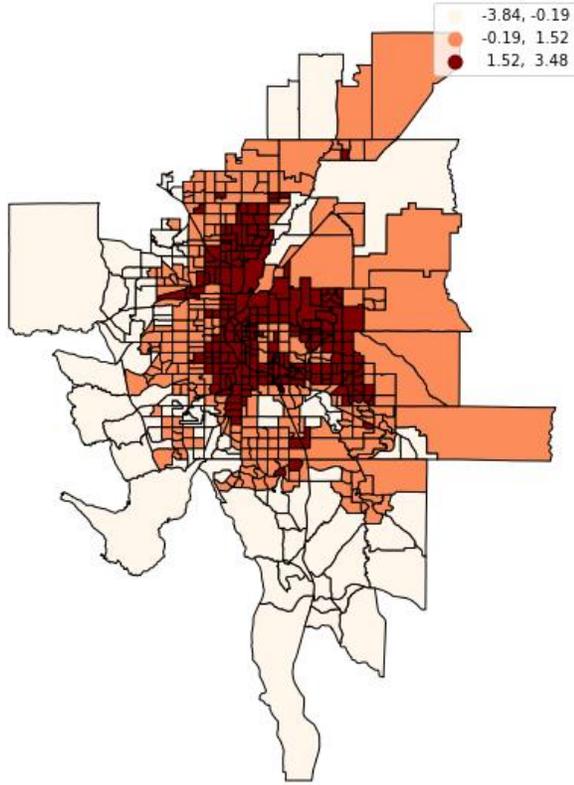 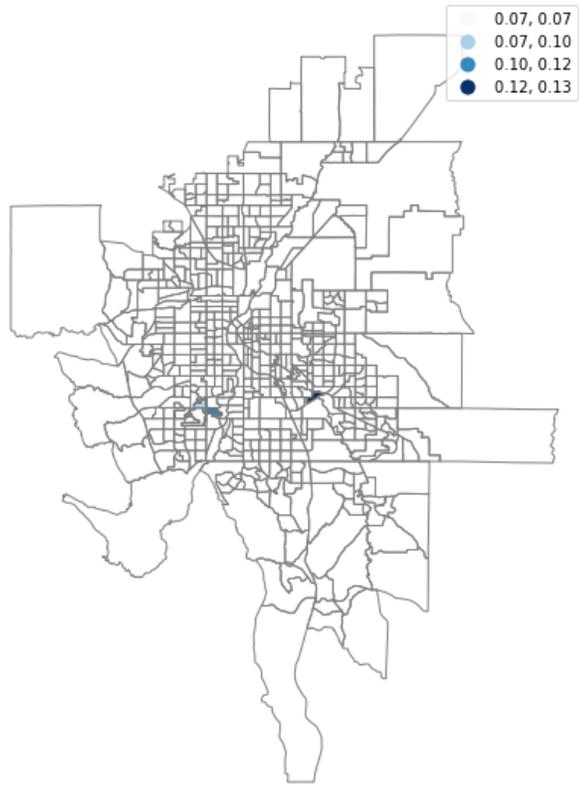

(A) Distribution of urban heat  (B) The ratio of trips from low UHI to high UHI

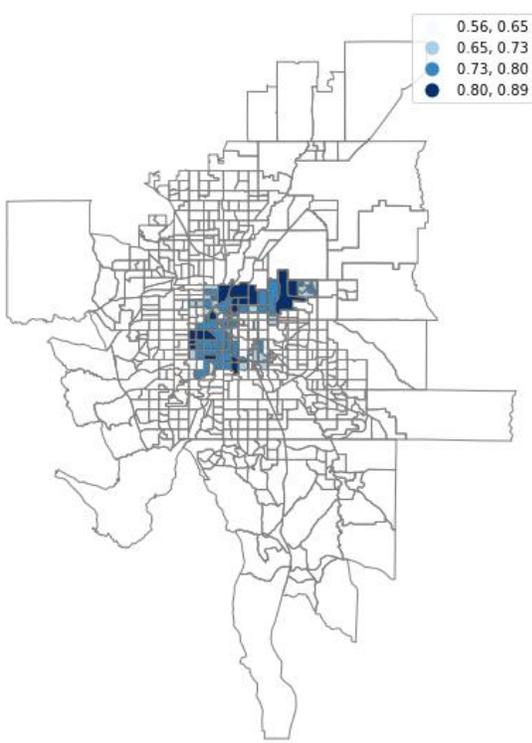 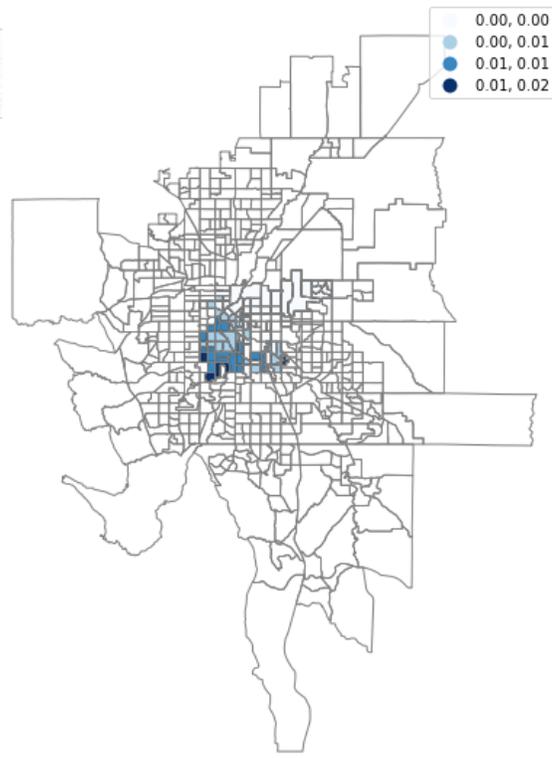

(C) The ratio of trips from high UHI to high UHI  (D) The ratio of trips from high UHI to low UHI

**Figure A - 14**. Urban Heat Traps and Trips in Denver Area. (A) shows that 17 percent of census tracts are low UHI areas, and 38 percent are in high UHI areas across Denver. (C) 44 percent of census tracts in high urban heat areas have trips to high urban heat census tract, representing that Denver is a metropolitan area with low urban heat traps.

**Appendix B. Urban Centrality Index, Income, White, and Non-white Gini indices**

**Table B.** Urban Centrality index (UCI), Spatial distribution of urban heat index, Income, White, and None-white Gini indices in each metropolitan area.

| MSA | UHI Spatial Gini | UCI | Income Gini | White Gini | Non-white Gini |
|---|---|---|---|---|---|
| Atlanta, GA | 0.76 | 0.49 | 0.47 | 0.71 | 0.72 |
| Boston, MA | 0.59 | 0.32 | 0.48 | 0.59 | 0.53 |
| Chicago, IL | 0.56 | 0.42 | 0.48 | 0.66 | 0.65 |
| Columbus, OH | 0.50 | 0.56 | 0.46 | 0.54 | 0.58 |
| Dallas. TX | 0.56 | 0.51 | 0.47 | 0.50 | 0.45 |
| Denver, CO | 0.71 | 0.75 | 0.45 | 0.50 | 0.49 |
| Detroit, MI | 0.48 | 0.40 | 0.47 | 0.76 | 0.80 |

| City | | | | | |
|---|---|---|---|---|---|
| Houston, TX | 0.39 | 0.50 | 0.48 | 0.50 | 0.46 |
| Los Angeles, CA | 0.41 | 0.42 | 0.49 | 0.55 | 0.48 |
| Memphis, TN | 0.92 | 0.64 | 0.50 | 0.10 | 0.22 |
| Miami, FL | 0.55 | 0.41 | 0.51 | 0.48 | 0.58 |
| Minneapolis, MN | 0.62 | 0.57 | 0.44 | 0.50 | 0.54 |
| Orlando, FL | 0.99 | 0.53 | 0.47 | 0.43 | 0.44 |
| Philadelphia, PA | 0.56 | 0.35 | 0.48 | 0.71 | 0.68 |
| Phoenix, AZ | 0.85 | 0.72 | 0.39 | 0.48 | 0.47 |
| Pittsburgh, PA | 0.80 | 0.47 | 0.48 | 0.53 | 0.61 |
| Portland, OR | 0.54 | 0.71 | 0.45 | 0.29 | 0.37 |
| Rochester, NY | 0.77 | 0.47 | 0.46 | 0.60 | 0.65 |
| Seattle, WA | 0.82 | 0.55 | 0.47 | 0.39 | 0.43 |
| Washington DC | 0.52 | 0.43 | 0.45 | 0.66 | 0.68 |

UHI Gini index is calculated based on the average UHI indices in each metropolitan area. The higher the UHI Gini index, the more clustered UHI areas. UCI, also known as the urban centrality index, assesses the centrality of a certain area (city, metropolitan area, region, country, etc.) on a continuum ranging from extreme monocentric to extreme polycentric (Pereira et al., 2013). UCI values vary from 0 to 1, with 0 expressing the highest level of polycentricity and 1 the highest level of monocentricity. Income, White, and non-White Gini indices are retrieved from the American Community Survey database administrated by US Census Bureau ("United States Census Bureau,") 5-year data. These Gini indices vary from 0 to 1, with 0 representing perfectly not segregated neighborhoods and 1 representing perfectly segregated neighborhoods.

**Appendix C: Statistical Significance**

**Table C.** Statistical Significance of traps escalates and escapes vs. Urban Centrality Index, Spatial Distribution of Urban Heat Index, and Income, White, and non-white Gini indices. There is no statistical significance between traps and demographic segregation.

|          | UHI Spatial Gini | UCI   | Income Gini | White Gini | Non-White Gini |
|----------|------------------|-------|-------------|------------|----------------|
| trap     | 0.01             | -0.24 | 0.37        | -0.09      | 0.01           |
| escalade | 0.15             | -0.11 | 0.05        | -0.08      | 0.03           |
| escape   | 0.08             | 0.1   | -0.14       | 0.17       | 0.12           |